\documentclass[12pt,aps,prd,superscriptaddress,groupedaddress,floatfix,nofootinbib]{revtex4-2}
\pdfoutput=1
\usepackage[dvipsnames]{xcolor}
\usepackage{dcolumn}
\usepackage{bm}
\usepackage{graphicx}
\usepackage{amssymb,amsmath}
\usepackage{multirow}
\usepackage{units,changes}
\usepackage{url}
\usepackage{mathtools}
\usepackage{tabu}
\usepackage{array}
\usepackage[colorlinks=true,urlcolor=blue,anchorcolor=blue
,citecolor=blue,filecolor=blue,linkcolor=blue,menucolor=blue
,linktocpage=true,pdfproducer=medialab,pdfa=true]{hyperref}
\usepackage{slashed}
\DeclareUnicodeCharacter{2212}{-}
\graphicspath{{Figs/}}
\def\beqa{\begin{eqnarray}}
\def\eeqa{\end{eqnarray}}

\def\figureautorefname~#1\null{Fig.\,#1\null}
\def\tableautorefname~#1\null{Tab.\,#1\null}

\def\equationautorefname~#1\null{Eq.\,(#1)\null}

\begin{document}

\title{Probing long-lived doubly charged scalar in the Georgi-Machacek model at the LHC and in far detectors}

\author{Chih-Ting Lu}
\email{ctlu@njnu.edu.cn}
\affiliation{Department of Physics and Institute of Theoretical Physics, Nanjing Normal University, Nanjing, 210023, China}

\author{Xinyu Wang}
\email{xinyuwang@njnu.edu.cn}
\affiliation{Department of Physics and Institute of Theoretical Physics, Nanjing Normal University, Nanjing, 210023, China}

\author{Xinqi Wei}
\email{xqwei@njnu.edu.cn}
\affiliation{Department of Physics and Institute of Theoretical Physics, Nanjing Normal University, Nanjing, 210023, China}

\author{Yongcheng Wu}
\email{ycwu@njnu.edu.cn}
\affiliation{Department of Physics and Institute of Theoretical Physics, Nanjing Normal University, Nanjing, 210023, China}

\date{\today}

\begin{abstract}
Searching for long-lived particles (LLPs) beyond the Standard Model (SM) is a promising direction in collider experiments. The Georgi-Machacek (GM) model extends the scalar sector in the SM by introducing various new scalar bosons. In this study, we focus on the parameter space that allows the light doubly charged scalar to become long-lived. This light doubly charged scalar is fermophobic and predominantly decays into a pair of on-shell or off-shell same-sign $W$ bosons. We investigate three types of signal signatures at the LHC: displaced vertices in the inner tracking detector, displaced showers in the muon system, and heavy stable charged particles. Additionally, we analyze the potential for detecting such doubly charged scalars in far detectors, including ANUBIS, MATHUSLA, FACET, FASER, CODEX-b, MoEDAL-MAPP and AL3X. By combining the LLP searches at the LHC and in far detectors, we project that the limits on the mixing angle, $\theta_H$, (between the doublet and triplets) can cover most of the parameter space with $\sin\theta_H\lesssim 10^{-3}$ for the mass range of long-lived doubly charged scalars between $50$ GeV to $180$ GeV, assuming the full integrated luminosity at the LHC and HL-LHC.
\end{abstract}

\maketitle
\newpage
\tableofcontents
\newpage

\section{Introduction}

Searching for new scalar bosons beyond the Standard Model (SM) is an important mission at the Large Hadron Collider (LHC) and future colliders. Some simple extensions of the scalar sector in the SM include scalar singlet models~\cite{Robens:2015gla,Wang:2023suf}, two-Higgs doublet models (2HDM)~\cite{Lee:1973iz,Branco:2011iw,Wang:2022yhm}, scalar triplet models~\cite{Gunion:1989ci,Englert:2013zpa}, and the Georgi-Machacek (GM) model~\cite{Georgi:1985nv,Chanowitz:1985ug,chiang2014novel,chiang2019global,du2023explaining,du2024positive,sun2017searching,Das:2018vkv,Chakraborti:2023mya,Kundu:2021pcg,Chowdhury:2024mfu,Bairi:2022adc,Ahriche:2022aoj}, among others. The GM model is particularly notable for providing rich phenomenological predictions at collider experiments~\cite{ATLAS:2022zuc,CMS:2021wlt,CMS:2019qfk,ATLAS:2017uhp,ATLAS:2015edr,Logan:2018wtm}, neutrino experiments~\cite{Richard:2021ovc}, and gravitational wave experiments~\cite{Chen:2022zsh,Bian:2019bsn}. Moreover, the GM model can explain some experimental anomalies, including the excess of $95$ GeV signal in $\gamma \gamma$, $b \Bar{b}$, $\tau \tau$ channels~\cite{Ahriche:2023wkj}. In the GM model, the CP-even singlet as well as the triplet and fiveplet scalars can all be used as candidates for the mass with $95$ GeV to explain this possible excess.

The GM model introduces extra complex and real triplets in addition to the SM particles, which ensures the preservation of custodial symmetry at tree level. After electroweak symmetry breaking (EWSB), the GM model includes a fiveplet, two triplets, and two singlets where one of the singlets will be identified as the SM-like Higgs boson with the mass of 125 GeV, one of the triplets corresponds to the goldstone.
The extra triplet and singlet are similar to those in the 2HDM, which have been searched for in many channels~\cite{ATLAS:2024rzd,ATLAS:2024jja,ATLAS:2024vxm,ATLAS:2024lyh,ATLAS:2023zkt,ATLAS:2022fpx,ATLAS:2019nkf,ATLAS:2019tpq,CMS:2023xpx,CMS:2019bnu,CMS:2018uag}.
For the fiveplet, the searches mainly focus on the charged components especially the doubly charged component. In general, several channels have been probed for a general doubly charged scalar, including the Drell-Yan production~\cite{ATLAS:2021jol}, vector-boson-fusion (VBF) process~\cite{ATLAS:2020ius,CMS:2021wlt,ATLAS:2024txt} and those involving lepton number violation at the LHC~\cite{CMS:2012dun,ATLAS:2022pbd,altakach2022discovery,mitsou2024looking} as well as the production from $ep$ collision at HERA~\cite{H1:2006ecx} and $e^+e^-$ collision at LEP~\cite{L3:2003zst, OPAL:2003kya, DELPHI:2002bkf, OPAL:2001luy}. However, the searches involving the lepton number violation interaction are not applicable to the doubly charged component of the fiveplet in GM model, as the fiveplet is fermiophobic at tree-level.
Hence the most relevant ones are those where the doubly charged scalar is produced and decaying through its coupling with gauge bosons. For the same reason, the single charged fivplet is also mainly searched through its coupling with gauge boson utilizing the VBF production and associated production with other fiveple and the decay channels to gauge bosons~\cite{CMS:2019qfk,ATLAS:2015edr,ATLAS:2021jol,ATLAS:2022zuc}.
The coupling between the fiveplet and the gauge bosons is proportional to the triplet vacuum expectation value (vev). In general, suppression of the contribution from the triplet vev to the total one will not affect those analysis based on its gauge couplings. However, when the triplet vev is highly suppressed, as well as when the mass of fiveplet is below the decay threshold of two on-shell gauge bosons, the decay length of $H_5^{\pm\pm}$ will increase, it may become long-lived particle (LLP) which alters the analysis based on the gauge boson final states. The current analysis hence loss its sensitivity in such low mass and low triplet vev region. In this work, we will focus on the parameter region where $H_5^{\pm\pm}$ is long-lived.

Due to their distinct experimental signatures, LLPs play a crucial role in the search for new physics beyond the SM. LLPs can travel significant distances within a detector before decaying, producing observable phenomena such as displaced vertices or unusual energy deposits that may be missed by traditional searches focused on prompt decays. These unique characteristics make LLPs invaluable for probing new physics, particularly in parameter regions that are otherwise difficult to access. In the GM model, the $H_5^{\pm\pm}$ can become an LLP under certain conditions, such as when the triplet vev is highly suppressed or when its mass is below the threshold for decays into two on-shell same-sign $W$ bosons. Studying LLPs enhances the experimental sensitivity at the LHC, enabling exploration of new parameter spaces and providing deeper insights into the mechanisms of electroweak symmetry breaking and the structure of scalar sectors.

The main focus of this work is the search for the long-lived $H_{5}^{\pm\pm}$ in the GM model at the LHC and far detectors\footnote{There are some previous studies for the long-lived doubly charged scalars in Type-II seesaw model and Left-right symmetry model at the LHC~\cite{Han:2007bk,BhupalDev:2018tox,Antusch:2018svb,Akhmedov:2024rvp}.}. If the charged LLPs have lifetimes sufficient to travel detectable distances within the detector before decaying, several general searches can be conducted at the LHC and in far detectors:
\begin{itemize}
\item Inner tracking detector:
LLPs travel a measurable distance before decaying, leading to displaced vertices within the inner tracking detector away from the primary collision point~\cite{CMS:2014hka}. LLPs decaying into charged particles can leave tracks with large impact parameters relative to the primary vertex. Searches for events with such high-impact parameter tracks are carried out~\cite{Araz:2021akd}.
\item Muon system: Some LLPs can traverse the electromagnetic calorimeter, hadronic calorimeter and decay into leptons and/or jets in the muon system, where they are reconstructed as a cluster-type signal, leaving distinctive signatures in the muon system. Searches for excesses of energetic clusters are conducted~\cite{CMS:2021juv}.
\item Tracking detectors: Some LLPs live long enough to fly out of the detectors, leaving a unique charged track signature inside the detectors, hence the name heavy stable charged particles (HSCPs).
\item Far detectors: Near each collision point, several far detectors are under construction or in preparation, such as ANUBIS~\cite{Bauer:2019vqk},  MATHUSLA~\cite{MATHUSLA:2020uve}, FACET~\cite{Cerci:2021nlb}, FASER, FASER2~\cite{FASER:2018eoc}, CODEX-b~\cite{Gligorov:2017nwh}, MoEDAL-MAPP~\cite{Pinfold:2019nqj} and AL3X~\cite{Gligorov:2018vkc}. Some LLPs have a sufficiently long lifetime to fly out of ATLAS/CMS/LHCb/ALICE detectors and reach these far detectors.
\end{itemize}
Therefore, we focus on the following four detector areas to search for long-lived $H_{5}^{\pm\pm}$: (1) Inner tracker system; (2) Muon system; (3) Heavy stable charged particles (HSCPs); (4) Far detectors (ANUBIS, MATHUSLA, FACET, FASER, CODEX-b, MoEDAL-MAPP and AL3X).

The rest of this paper is organized as follows.
We briefly review the GM model and the production and decay of the doubly charged scalar in Sec.~\ref{Sec:GM_model}, from which we will demonstrate the parameter region for long-lived $H_5^{\pm\pm}$.
The search strategies and results for the long-lived doubly charged scalar at the LHC and in the far detectors are presented in Sec.~\ref{Sec:LHC_LLPs}.
Finally, we present our findings and discussions in Sec.~\ref{Sec:Conclusion}.

\section{The Georgi-Machacek model}
\label{Sec:GM_model}
\subsection{Model setup and low-mass benchmark}
The scalar sector of GM model~\cite{Georgi:1985nv,Chanowitz:1985ug} includes the usual Higgs doublet $\phi=( \phi^{\dagger},\phi^{0} )^{T}$, and extra two triplets, one complex $\chi=( \chi^{++}, \chi^{+}, \chi^{0})^{T}$ and one real $\xi=( \xi^{+}, \xi^{0}, -\xi^{+*})^{T}$. The fields are arranged in the form of bi-doublet and bi-triplet in order to make the global symmetry ${\rm SU}(2)_{\rm L}\times{\rm SU}(2)_{\rm R}$ manifest:
\begin{align}
\Phi &\equiv (\varepsilon_{2} \phi^{*},\phi)=
\begin{pmatrix}
\phi^{0*}&\phi^{+}\\
-\phi^{+*}&\phi^{0}
\end{pmatrix},\quad
\rm{with} \quad \varepsilon_{2}=
\begin{pmatrix}
0&1\\
-1&0
\end{pmatrix},\\
X &\equiv ({\varepsilon_{3} \chi^{*},\xi,\chi})=
\begin{pmatrix}
\chi^{0*}&\xi^{+}&\chi^{++}\\
-\chi^{+*}&\xi^{0}&\chi^{+}\\
\chi^{++*}&-\xi^{+*}&\chi^{0}
\end{pmatrix},\quad
\rm{with} \quad \varepsilon_{3}=
\begin{pmatrix}
0&0&1\\
0&-1&0\\
1&0&0
\end{pmatrix}.
\end{align}
The most general gauge invariant scalar potential involving $\Phi$ and $X$ that is also preserving the global ${\rm SU}(2)_{\rm L}\times{\rm SU}(2)_{\rm R}$ can be written as~\cite{Hartling:2014zca}
\begin{align}
V(\Phi,X)&=\frac{\mu_{2}^{2}}{2} {\rm Tr}(\Phi^\dagger \Phi)+\frac{\mu_{3}^{2}}{2} {\rm Tr}(X^\dagger X)+\lambda_{1}\left[{\rm Tr}(\Phi^{\dagger} \Phi)\right]^2+\lambda_2 {\rm Tr}(\Phi^\dagger\Phi){\rm Tr}(X^\dagger X)\nonumber \\
&\quad+\lambda_3 {\rm Tr}(X^\dagger X X^\dagger X)+\lambda_{4}\left[ {\rm Tr}(X^\dagger X)\right]^{2}-\lambda_{5}{\rm Tr}(\Phi^\dagger \tau^a \Phi \tau^b){\rm Tr}(X^\dagger t^a X t^b)\nonumber\\
&\quad-M_1 {\rm Tr}(\Phi^\dagger\tau^a\Phi\tau^b)(UXU^\dagger)_{ab}-M_2 {\rm Tr}(X^\dagger t^a X t^b)(UXU^\dagger)_{ab},
\end{align}
where the $\tau^{i}$ and $t^{i}$ ($i=1,2,3$) correspond to the $SU(2)$ generators for the doublet and triplet representations, respectively. $\tau^{i}=\sigma^{i}/{2}$ where $\sigma^{i}$ are the Pauli matrices and $t^{i}$'s are given by
 \begin{align}
t^1=\frac{1}{\sqrt{2}}
\begin{pmatrix}
0&1&0\\
1&0&1\\
0&1&0
\end{pmatrix},\quad
t^2=\frac{1}{\sqrt{2}}\begin{pmatrix}
0&-i&0\\
i&0&-i\\
0&i&0
\end{pmatrix},\quad
t^3=\begin{pmatrix}
1&0&0\\
0&0&0\\
0&0&-1
\end{pmatrix}.
\end{align}
The matrix $U$ is given by~\cite{Aoki:2007ah}
\begin{align}
U=\begin{pmatrix}
\frac{1}{\sqrt{2}}&0&\frac{1}{\sqrt{2}}\\
-\frac{i}{\sqrt{2}}&0&-\frac{i}{\sqrt{2}}\\
0&1&0
\end{pmatrix}.
\end{align}
Hence, in total, from the scalar potential, we have 9 free parameters of which $\mu_2^2$, $\mu_3^2$, $M_1$ and $M_2$ are dimensional and $\lambda_{1,\cdots,5}$ are dimensionless.

The vacuum expectation values (vevs) of the bi-doublet $\Phi$ and the bi-triplet $X$ are defined as
\begin{align}
\langle \Phi \rangle =\frac{v_{\phi}}{\sqrt{2}}I_{2 \times 2},\qquad \langle X \rangle =v_{\chi} I_{3 \times 3},
\end{align}
where the vevs from both doublet and triplets contribute to the gauge boson mass giving rise to the constraint
\begin{align}
v_{\phi}^{2}+8v_{\chi}^{2} \equiv v^{2} =\frac{1}{\sqrt{2}G_{F}}\approx (246\,{\rm GeV})^{2}.
\end{align}
The real triplet $\xi$ and the complex triplet $\chi$ are assumed to obtain the same vev, ie., $v_{\xi}=v_{\chi}$, to maintain the custodial ${\rm SU}(2)_{C}$ symmetry at tree-level. Further, the neutral fields are expanded around the corresponding vevs and decomposed into real and imaginary components according to
\begin{align}
\phi^{0}=\frac{v_{\phi}}{\sqrt{2}}+\frac{\phi^{0,r}+i\phi^{0,i}}{\sqrt{2}},\quad \chi^{0}=v_{\chi}+\frac{\chi^{0,r}+i\chi^{0,i}}{\sqrt{2}},\quad \xi^{0}=v_{\chi}+\xi^{0}.
\end{align}
The vevs of the doublet and triplets should minimize the potential by definition which provides the following two minimization conditions:
\begin{align}
\frac{\partial V}{\partial v_\phi}&=\left (\mu_{2}^{2}+4\lambda_{1}v_{\phi}^{2}+6\lambda_{2}v_{\chi}^{2}-3\lambda_{5}v_{\chi}^{2}-\frac{3}{2}M_{1}v_{\chi}\right )v_{\phi}=0,\\
\frac{\partial V}{\partial v_\chi}&=3\mu_{3}^{2}v_{\chi}+6\lambda_{2}v_{\phi}^{2}v_{\chi}+12\lambda_{3}v_{\chi}^{3}+36\lambda_{4}v_{\chi}^{3}-3\lambda_{5}v_{\phi}^{2}v_{\chi}-\frac{3}{4}M_{1}v_{\phi}^{2}-18M_{2}v_{\chi}^{2}=0,
\end{align}
which will be used to replace $\mu_2^2$ and $\mu_3^2$ in terms of $v_\phi$ and $v_\chi$ as
\begin{align}
\mu_2^2 &= \frac{3}{2}M_1v_\chi - 4\lambda_1 v_\phi^2 - 3(2\lambda_2-\lambda_5)v_\chi^2, \\
\mu_3^2 &= \frac{M_1v_\phi^2}{4v_\chi} + 6M_2v_\chi - 4(\lambda_3 + 3\lambda_4)v_\chi^2 - (2\lambda_2-\lambda_5)v_\phi^2.
\end{align}

The doublet and two triplets provide in total 13 physical fields. After electroweak symmetry breaking (EWSB), three of the fields correspond to the goldstone bosons, the rest are real physical fields which are arranged according to the representation under the custodial symmetry ${\rm SU}(2)_C$ as two singlets, one triplet and one fiveplet. In terms of the original fields, the goldstone bosons are given by
\begin{align}
    G^\pm &= c_H \phi^\pm + s_H \frac{\chi^\pm+\xi^\pm}{\sqrt{2}} \nonumber \\
    G^0 &= c_H \phi^{0,i} + s_H \chi^{0,i}
\end{align}
where we have defined
\begin{align}
    c_H \equiv \cos\theta_H = \frac{v_\phi}{v},\quad s_H\equiv \sin\theta_H = \frac{2\sqrt{2}v_\chi}{v}.
\end{align}
The physical fiveplet and triplet are given by
\begin{align}
\begin{cases}
    H_{5}^{\pm\pm}= \chi^{\pm\pm}, \\
    H_{5}^{\pm}=\frac{\chi^{\pm}-\xi^{\pm}}{\sqrt{2}}, \\
    H_{5}^{0}=\sqrt{\frac{2}{3}} \xi^{0}-\sqrt{\frac{1}{3}} \chi^{0,r},
\end{cases}\qquad
\begin{cases}
H_{3}^{\pm}=-s_{H} \phi^{\pm}+c_{H}\frac{\chi^{\pm}+\xi^{\pm}}{\sqrt{2}}, \\
H_{3}^{0}=-s_{H}\phi^{0,i}+c_{H}\chi^{0,i}.
\end{cases}
\end{align}
The masses of the fiveplet and triplet are given by, in terms of the quartic couplings in the potential and the vevs,
\begin{align}
\label{equ:m5}
m_{5}^{2}&=8\lambda_{3}v_{\chi}^{2}+\frac{3}{2}\lambda_{5}v_{\phi}^{2}+\frac{M_{1}}{4v_{\chi}}v_{\phi}^{2}+12M_{2}v_{\chi}, \\
\label{equ:m3}
m_{3}^{2}&=\left( \frac{M_{1}}{4v_{\chi}}+\frac{\lambda_{5}}{2}\right)v^{2}.
\end{align}
The two singlets in the gauge basis are given by
\begin{align}
H_{1}^{0}&=\phi^{0,r}, \\
H_{1}^{0'}&=\sqrt{\frac{1}{3}}\xi^{0}+\sqrt{\frac{2}{3}}\chi^{0,r}.
\end{align}
The mass matrix in the gauge basis $(H_1^0, H_1^{0'})$ is given by
\begin{align}
\label{equ:mass_singlet}
\mathcal{M}^2 = \begin{pmatrix}
    \mathcal{M}_{11}^2 & \mathcal{M}_{12}^2 \\
    \mathcal{M}_{12}^2 & \mathcal{M}_{22}^2
\end{pmatrix}
\end{align}
with
\begin{align}
    \mathcal{M}_{11}^2 &= 8\lambda_1v_\phi^2, \\
    \mathcal{M}_{12}^2 &= \frac{\sqrt{3}}{2}\left(4(2\lambda_2 - \lambda_5)v_\chi - M_1\right)v_\phi, \\
    \mathcal{M}_{22}^2 &= \frac{M_1v_\phi^2}{4v_\chi} - 6M_2v_\chi + 8(\lambda_3 + 3\lambda_4)v_\chi^2.
\end{align}
These two singlets will thus further mix with each other to provide the mass eigenstates
\begin{align}
h&=\cos\alpha H_{1}^{0}-\sin\alpha H_{1}^{0'}, \\
H&=\sin\alpha H_{1}^{0}+\cos\alpha H_{1}^{0'}.
\end{align}
The corresponding masses are $m_h$ and $m_H$ where we always assume that $h$ is the SM-like Higgs. The mixing angle $\alpha$ is hence uniquely determined by the mass matrix in~\autoref{equ:mass_singlet} with the above assumption.
In the following analysis, we will always using $m_h=125\,\rm GeV$ as one of the input parameters to facilitate the matching with current measurements at the LHC. We hence focus on the low-$m_5$ as well as low-$s_H$ region for $H_5^{\pm\pm}$.

In GM model, the search strategies for the extra singlet and triplet are similar to that for the extra heavy scalars in 2HDM with corresponding scaling according to the couplings with SM gauge bosons and fermions. On the other hand, the fiveplet in GM model is fermiophobic, it has no couplings with fermions as it contains only the ${\rm SU}(2)_L\times{\rm SU}(2)_R$ triplet components~\footnote{However, in Type-II seesaw, the scalar triplet can have Yukawa couplings with the leptons~\cite{BhupalDev:2013xol}. In this analysis, we don't consider such scenario.}. The coupling of the fiveplet with gauge boson at tree-level is proportional to $s_H=2\sqrt{2}v_\chi/v$.
An important search strategy for the fiveplet focuses primarily on the doubly charged components, produced via Drell-Yan and vector boson fusion processes, and their subsequent decay into gauge bosons~\cite{ATLAS:2021jol,ATLAS:2024txt}\footnote{A deviation from the SM has been observed for a resonant $H_5^{\pm\pm}$ mass near 375 GeV, with a local (global) significance of 3.3 (2.5) standard deviations, as reported in Ref.~\cite{ATLAS:2024txt}. Further data is required to verify this excess.}.
Note that, for the general doubly charged Higgs, there are actually two important decay modes, one is decaying into leptons for low triplet vev, and the other is decaying into W boson pair for high triplet vev~\cite{Muhlleitner:2003me,FileviezPerez:2008wbg}. However due to its fermiophobic nature, those only involving couplings with gauge bosons are the relevant ones.
Current LHC searches focus on the region where $m_5\geq200\,\rm GeV$ and can cover the parameter space up to $m_5\lesssim 350\,\rm GeV$. However, these search do not extend to the lower mass region. Due to the large Drell-Yan pair production cross section in the low mass region, which does not scale with $s_H$, and the unique decay channel $H_5^{\pm\pm}\to W^\pm W^\pm$ (assuming other scalars are heavy, a natural assumption under current LHC searches), one might expect that this channel could easily cover the lower mass region. However, in this region, there is a special scenario where the above searches base on $H_5^{\pm\pm}\to W^\pm W^\pm$ could fail. If a small $s_H$ suppresses the corresponding decay width, $H_5^{\pm\pm}$ may become a LLP, and the signatures would be significantly different from the usual multi-gauge bosons searches.

\begin{table}[!tbp]
\centering
\begin{tabular}{|l|l|l|}
\hline\hline
Fixed SM Parameters & Variable Parameters & Other Parameters in the Potential\\
\hline
$G_F = 1.1663787\times10^{-5}\,{\rm GeV}^{-2}$ & $m_5\in(50,180)\,\rm GeV$ & $\lambda_3=-\lambda_4=-1.5$ \\
$m_h = 125\,\rm GeV$ & $s_H\in(10^{-9},10^{-2})$ &  $\lambda_5 = -4\lambda_2 = -0.32(m_5/100\,{\rm GeV})$ \\
& & $M_2 = 10\,\rm GeV$\\
\hline\hline
\end{tabular}
\caption{The low-$m_5$ benchmark for the GM model~\cite{Ismail:2020kqz}, with emphasis on low-$s_H$ region.}
\label{tab:lowm5_benchmark}
\end{table}

The most relevant parameters in the GM model for this scenario are the mass of the fiveplet $m_5$ which determines the kinematics and $s_H$ which determines the decay length $c\tau$ of $H_5^{\pm\pm}$. Other parameters are irrelevant at leading order as long as the triplet are heavier than the fiveplet. Hence, in general, we have freedom to choose other model parameters and only focus on the $m_5$-$s_H$ parameter space plane. Here, for convenient, we choose the low-$m_5$ benchmark from~\cite{Ismail:2020kqz} where the possible constraints in the $m_5$-$s_H$ plane are also discussed. Although some constraints become irrelevant when $H_5^{\pm\pm}$ becomes long-lived, the most significant constraint still arises from $H_5^0\to\gamma\gamma$, depending on the specifics of the parameter space, as we will discuss below. The choice of model parameters in low-$m_5$ benchmark are given in~\autoref{tab:lowm5_benchmark}, where we have an emphasize on the low-$s_H$ region by scanning in log-scale. Instead of using the parameters in the potential ($\mu_2^2,\mu_3^2,\lambda_{1,\cdots,5},M_1,M_2$), this benchmark uses ($v,s_H,\lambda_{2,\cdots,5},M_2,m_h,m_5$) where $\lambda_1$ is fixed by $m_h$ given the singlet masses relation, $M_1$ is fixed by $m_5$ according to~\autoref{equ:m5}. Note that, the whole parameter space defined by low-$m_5$ benchmark is also subjected to the theoretical constraints which is included through {\tt GMCALC}~\cite{Hartling:2014xma}. However, we emphasize again that the following analysis depends only on $m_5$ and $s_H$, the other parameters can be tuned as long as we have $m_5<m_3$. Before we go into detail analysis of the signatures from long-lived $H_5^{\pm\pm}$, we discuss its production and decay in the rest of this section.

\subsection{The production and decay of \texorpdfstring{$H_5^{\pm\pm}$}{H5pp}}
\subsubsection{The pair production of \texorpdfstring{$H_5^{\pm\pm}$}{H5pp}}

\begin{figure}[!btp]
\centering
\includegraphics[width=0.8\textwidth]{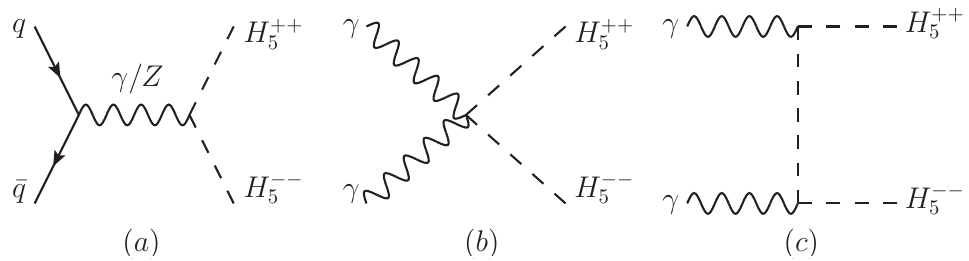}
\caption{The Feynman diagrams for the pair production of $H_5^{\pm\pm}$ including the Drell-Yan channel (a) and photon initiated channel (b, c and corresponding u-channel diagram).}
\label{fig:production_diagrams}
\end{figure}

\begin{figure}[!tbp]
\centering
\includegraphics[width=0.6\textwidth]{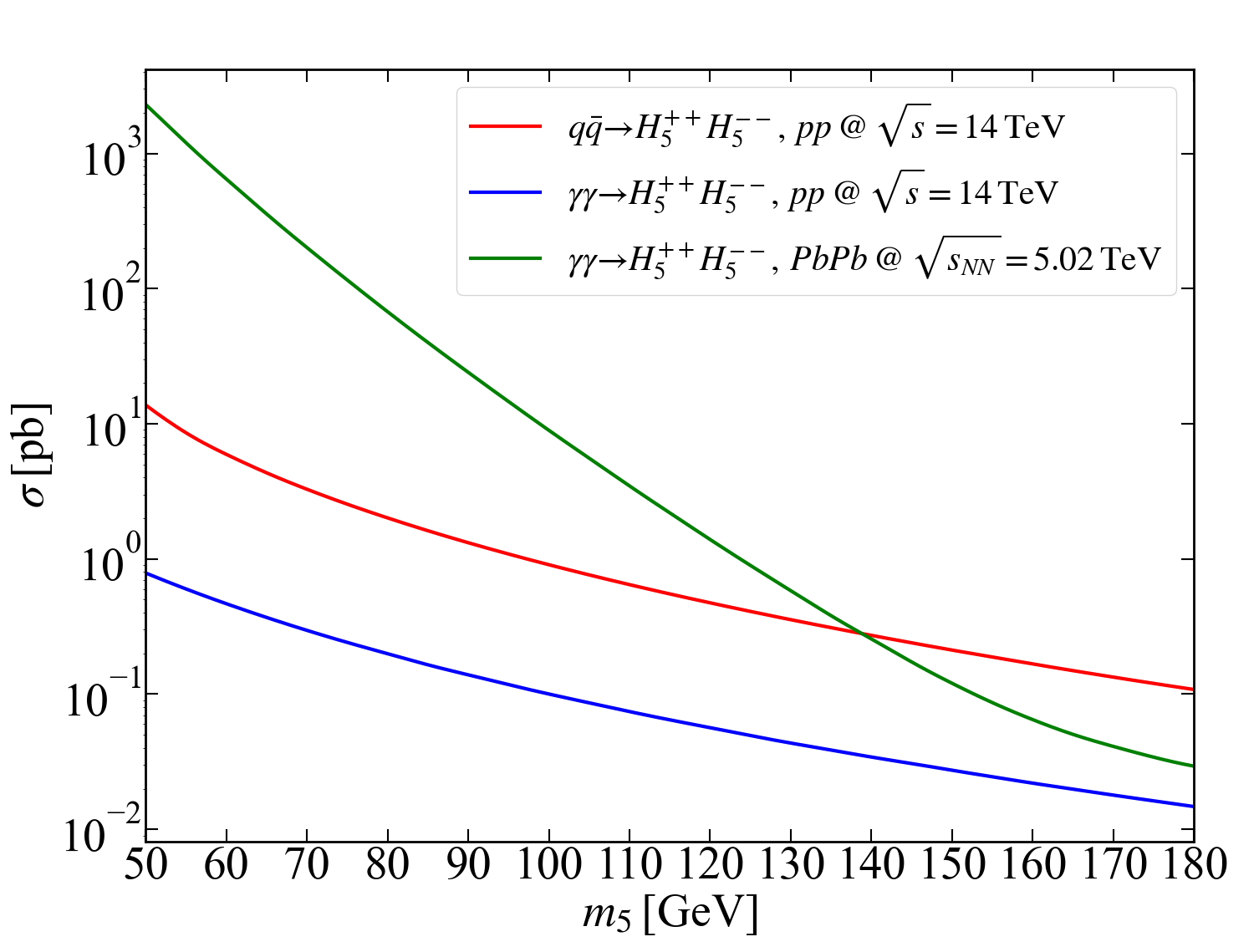}
\caption{The cross section for pair production of $H_5^{\pm\pm}$ as function of its mass $m_5$ including the Drell-Yan production channel (solid red) and photon initiated channel (solid blue) from p-p collision, as well as the photon initiated channel from Pb-Pb collision (solid green).}
\label{fig:m5_sigma}
\end{figure}

The production of $H_5^{\pm\pm}$ is dominated by Drell-Yan pair production at low-mass region followed by photon initiated channel. The diagrams for these channels are shown in~\autoref{fig:production_diagrams}. The couplings in GM model involved in the production are the couplings between $H_5^{\pm\pm}$ and the $\gamma/Z$ gauge bosons. The corresponding Feynman rules are given by
\begin{align}
    g_{H_5^{++}H_5^{--}\gamma} &= -2ie(p_{H_5^{++}}-p_{H_5^{--}})^\mu,\qquad g_{H_5^{++}H_5^{--}Z} = 2ie \frac{c_{2W}}{s_{2W}}(p_{H_5^{++}}-p_{H_5^{--}})^\mu,\nonumber\\
    g_{H_5^{++}H_5^{--}\gamma\gamma} &= 8ie^2g^{\mu\nu},
\end{align}
where it is important to notice that all these couplings are not suppressed by $s_H$. Hence, the cross section of these production channels do not depend on $s_H$. The cross sections of these production channels as functions of the mass are shown in~\autoref{fig:m5_sigma} for different beam configurations. The cross sections are obtained using {\tt MadGraph5}~\cite{Alwall:2011uj} with the UFO model file provided by~\cite{UFO:GM} and the parameter card generated from {\tt GMCALC}~\cite{Hartling:2014xma}. For the proton collision, {\tt NNPDF 3.1}~\cite{NNPDF:2017mvq,Bertone:2017bme} including {\tt LUXqed} formalism for the photon PDF~\cite{Manohar:2016nzj,Manohar:2017eqh} is used. While for heavy-ion collision, the photon flux in ultraperipheral collision incorporated in {\tt MadGraph5} are used~\cite{Shao:2022cly}. From proton collision, the Drell-Yan pair production is the dominant channel that can reach $\mathcal{O}(1)\,\rm pb$ at low-$m_5$ region. The photon initiated channel including both elastic and inelastic contribution is subdominant and provides $\lesssim 10\%$ corrections. From a simple estimation, the contribution from elastic photon production in heavy-ion collisions will be enhanced by $Z^4$, where $Z$ is the proton number of the ion. In~\autoref{fig:m5_sigma}, we also show the cross section of such contribution from Pb-Pb collision with $\sqrt{s_{NN}} = 5.02\,\rm TeV$. When the mass of $H_5^{\pm\pm}$ is small, the cross section is significantly enhanced. However, at higher mass regions, the cross section is greatly suppressed due to phase space limitations. Additionally, given the relatively low luminosity of Pb-Pb collision, this process is difficult to probe at the LHC. Note that the $H_5^{\pm\pm}$ can also be produced associated with single-charged component ($H_5^\pm$) of the fiveplet. However, the discussion on the phenomenology of $H_5^\pm$ depends on the details of the parameter space. In this work, we focus on the analysis of the signature from $H_5^{\pm\pm}$ with the dependence only on two parameters $m_5$ and $s_H$, which can thus be easily generalized to any other situation with doubly-charged scalar particles.

\subsubsection{The decay of \texorpdfstring{$H_5^{\pm\pm}\to W^\pm W^\pm$}{H5pp to Wp Wp}}
In the GM model, no lepton number violation interaction is introduced. Hence, if the fiveplet is lighter than the scalar triplet (which contains a single-charged component $H_3^\pm$), $H_5^{\pm\pm}$ has only one decay channel $H_5^{\pm\pm}\to W^\pm W^\pm$. Although, the branching ratio will never change, the decay width highly relies on the mass $m_5$ and the couplings between $H_5^{\pm\pm}$ and $W^\pm$ which is proportional to $s_H$~\cite{CMS:2017fhs}. In particular, when the mass $m_5$ is below the threshold $2m_W$, the decay can only happen through at least one off-shell $W$. On the other hand, current measurements from direct searches of heavy particles and from the Higgs signal strength give rise to strong constraints on $s_H$ which indicates the contribution of the ${\rm SU}(2)_L\times{\rm SU}(2)_R$ triplet vev to the total vev. Both situations will greatly suppress the decay width of $H_5^{\pm\pm}$ in the low-$m_5$ and low-$s_H$ region.

Analytically, the decay width of $H_5^{\pm\pm}\to W^{\pm}W^{\pm}$ including the off-shell effect is given by~\cite{Romao:1998sr,Contino:2014aaa}
\begin{align}
\label{equ:width_offshell}
\Gamma(H_{5}^{\pm\pm} \rightarrow W^{\pm}W^{\pm})=\frac{1}{\pi^2}\int_0^{m_{H_{5}}^2} dQ_{1}^{2} & \int_0^{(m_{H_{5}} - Q_{1})^2} dQ_{2}^2  \nonumber \\
\times
\frac{Q_{1}^{2} \Gamma_{W^{\pm}} / M_{W^{\pm}}}{(Q_{1}^{2}-M_{W^{\pm}}^{2})^{2}+M_{W^{\pm}}^2\Gamma_{W^{\pm}}^2}& \frac{Q_{2}^2\Gamma_{W^{\pm}}/M_{W^{\pm}}}{(Q_{2}^{2}-M_{W^{\pm}}^{2})^{2}+M_{W^{\pm}}^{2}\Gamma_{W^{\pm}}^{2}} \Gamma^{H_{5}^{\pm\pm} W^{\pm} W^{\pm}}(Q_{1}^{2},Q_{2}^{2}),
\end{align}
where $\Gamma_{W^{\pm}}$ is the width of $W$ boson and $Q_{i}^{2}$ is the square of the four-momentum of $W$ boson. $\Gamma^{H_{5}^{\pm\pm} W^{\pm} W^{\pm}}(Q_{1}^{2},Q_{2}^{2})$ is given by
\begin{align}
\label{equ:width_onshell}
\Gamma^{H_{5}^{\pm\pm} W^{\pm} W^{\pm}} (Q_{1}^{2}, Q_{2}^{2})
= S_{V} \frac{|g_{H_{5}^{\pm\pm} W^{\pm} W^{\pm}}|^{2} m_{H_{5}^{\pm\pm}}^3}{64 \pi Q_{1}^{2} Q_{2}^{2}} \left[ 1 - 2 k_{1} - 2 k_{2} + 10 k_{1} k_{2} + k_{1}^{2} + k_{2}^{2} \right]\lambda^{1/2}(k_{1}, k_{2}),
\end{align}
where $S_{V}=1/2$ is the symmetry factor,
$k_{i}=Q_{i}^{2}/m_{5}^{2}$ and $\lambda(x,y)$ is the kinematic function given by
\begin{align}
\lambda(x,y) = (1 - x - y)^{2} - 4xy.
\end{align}
The coupling involved in this decay is given by
\begin{align}
    g_{H_5^{\pm\pm}W^\mp W^\mp} = \sqrt{2}g^2vs_H.
\end{align}
Hence, the decay width is suppressed by mainly two factors. First, it will be suppressed by the tails of at least one of the Breit-Wigner resonants factors in~\autoref{equ:width_offshell} when $m_5 < 2m_W$. The width will further be suppressed by $s_H$ through $g_{H_5^{\pm\pm}W^\mp W^\mp}$ in~\autoref{equ:width_onshell}. It is thus nature to expect that the decay width will be suppressed strongly in low-$m_5$ and low-$s_H$ region such that $H_5^{\pm\pm}$ will travel some distance before it decays into SM particles which significantly alters the signatures at the LHC.

\begin{figure}
\centering
\includegraphics[width=0.7\textwidth]{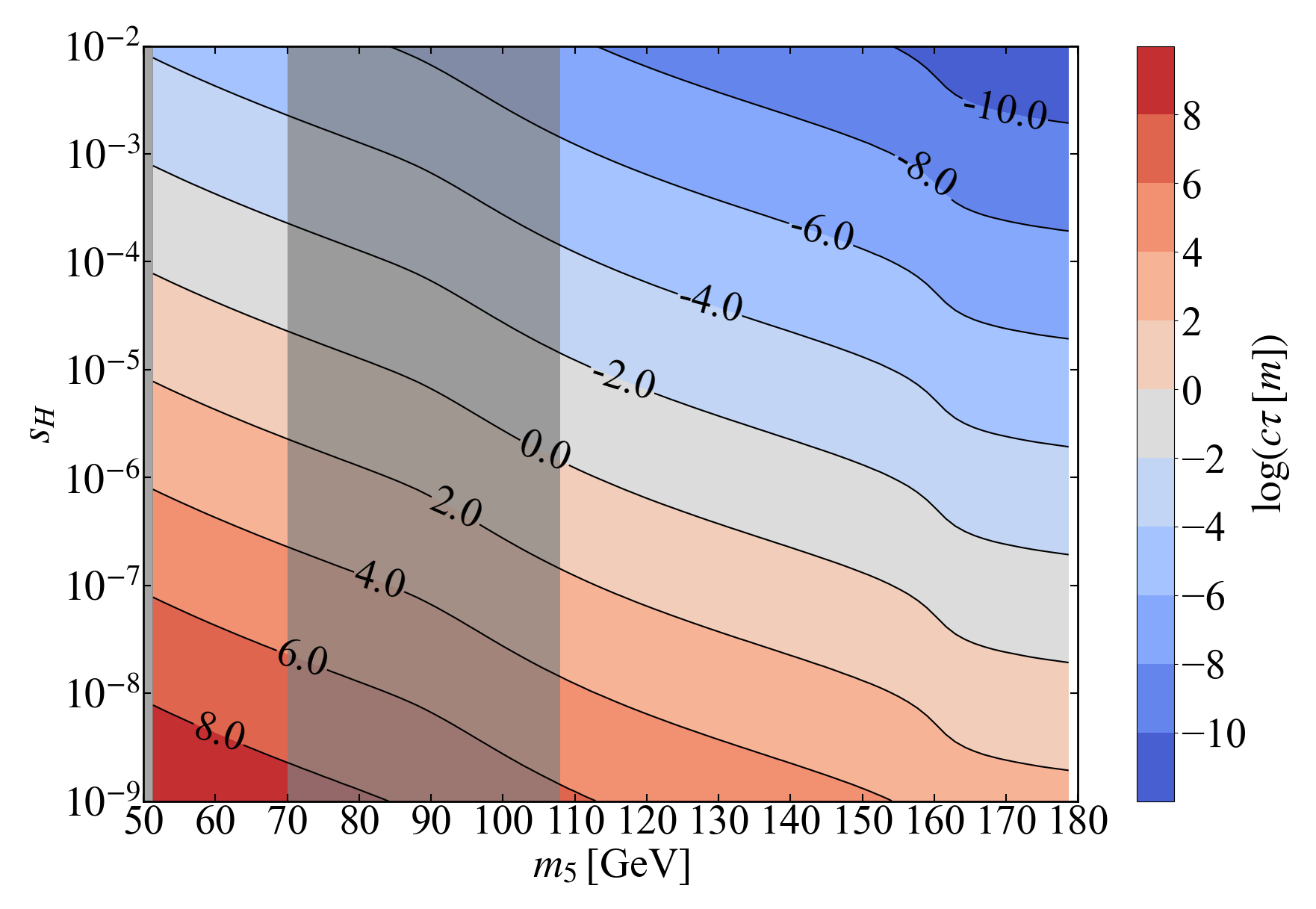}
\caption{The proper decay length ($c\tau$) of $H_{5}^{\pm\pm}$ in the $m_5$-$s_H$ plane. The gray areas indicate the parameter regions that are excluded by diphoton searches in low-$m_5$ benchmark.}
\label{fig:ctau_m5_sH}
\end{figure}

The proper decay length of $H_5^{\pm\pm}$ in the $m_5$-$s_H$ plane is shown in~\autoref{fig:ctau_m5_sH}. It is clear that the decay length $c\tau$ increase with the decrease of both $m_5$ and $s_H$. It shows clearly the threshold effects around $m_5\sim160\,\rm GeV$ as well. In the lower left corner, $c\tau$ can easily reach several meters or even hundred/thousand meters which is the parameter region that can be covered by LLP searches at the LHC as well as in the far detectors.
As we have discussed above, the long-lived nature of $H_5^{\pm\pm}$ in this parameter space invalidates most of the current searches of doubly-charged scalar. The most relevant search, which still depends on the details of the parameter space, comes from the diphoton resonant searches. The neutral component of the fiveplet $H_5^0$ will have a dominant decay channel $H_5^0\to\gamma\gamma$ in the mass region of interests, which is not suppressed by the phase space~\cite{Ismail:2020zoz}. Such channel has already been covered at the LHC from both ATLAS~\cite{ATLAS:2014jdv,ATLAS:2022abz} and CMS~\cite{CMS:2018cyk,CMS-PAS-HIG-13-001}. We recast the results in the low-$m_5$ benchmark and show the excluded region by gray area in~\autoref{fig:ctau_m5_sH}. The region with $70\,{\rm GeV}\lesssim m_5\lesssim 108\,\rm GeV$ and a tiny region with $m_5\sim 50\,\rm GeV$ of the low-$m_5$ benchmark are excluded by diphoton searches.
Note that the decay width of $H_5^0\to\gamma\gamma$ has contributions that do not depend on $s_H$, the diphoton searches can cover the relevant region extending the whole range of $s_H$. However, this depends on the details of the parameter space. In particular, different contributions i.e. from $W^\pm$ loop and $H_3^\pm$ loop, can cancel with each other to suppressed the decay ratio such that the diphoton searches no longer have sensitivity.

\section{Searching for the long-lived \texorpdfstring{$H_5^{\pm\pm}$}{H5pp} at the LHC and in far detectors}
\label{Sec:LHC_LLPs}
As discussed in previous section, in specific parameter region of $m_5$-$s_H$ plane, $H_5^{\pm\pm}$ becomes long-lived which significantly alters the signatures at the LHC. In this section, we will discuss the signatures of long-lived $H_5^{\pm\pm}$ at the ATLAS/CMS detectors as well as the detectors located far away from the interaction point (IP). In the first case, we will focus on the scenarios where $H_5^{\pm\pm}$ travels some distance before it decays within the detectors (in different layers) or $H_5^{\pm\pm}$ travels outside the detectors but still leaves signal through its charged track. Finally, we will discuss the case where $H_5^{\pm\pm}$ travels a long distance and leaves signals in various far detectors.

\begin{figure}[!b]
\centering
\includegraphics[width=0.95\textwidth]{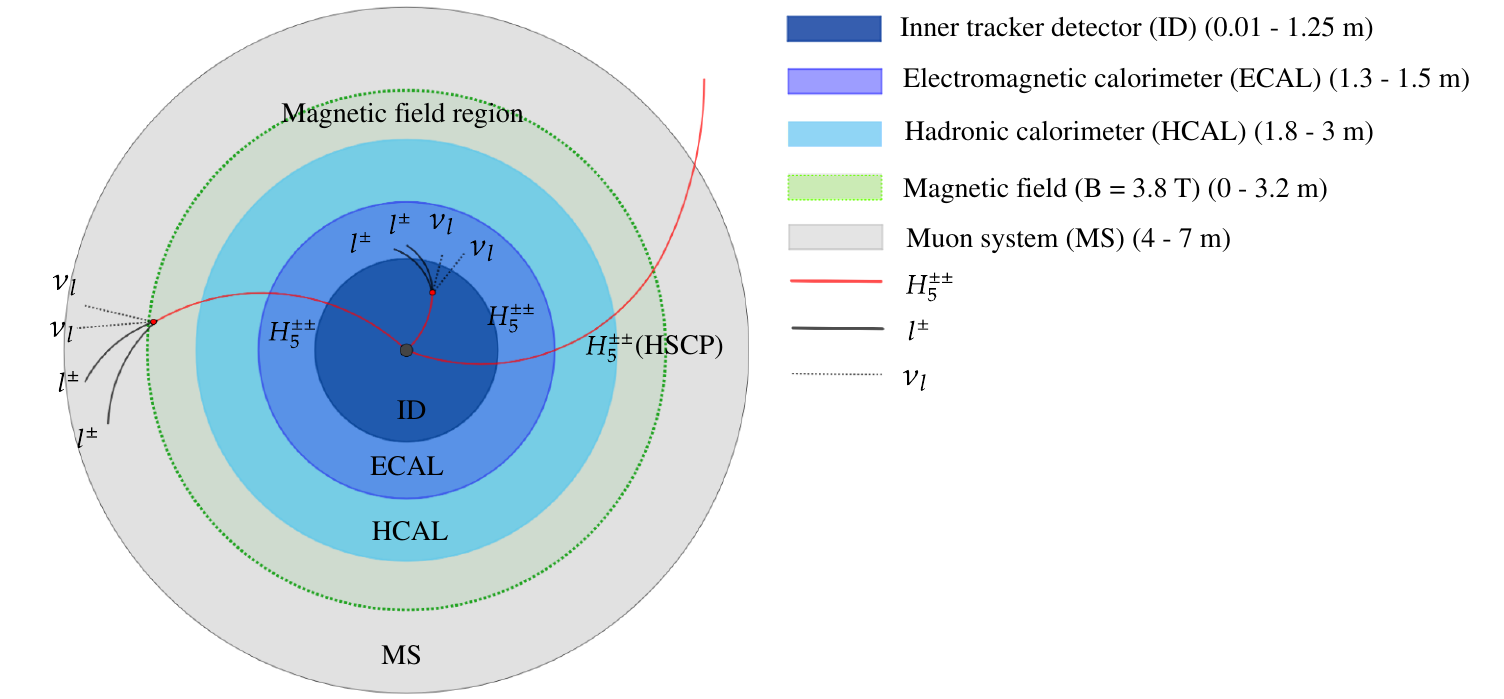}
\caption{Several main subsystems of the generic detectors (ATLAS/CMS) at the LHC~\cite{Lee:2018pag}.}
\label{fig:LHC_main_system}
\end{figure}

The general-purpose detectors, although are different in details, share similar structures as shown in~\autoref{fig:LHC_main_system}. From inside out, it consists of inner tracking detector (ID), electromagnetic calorimeter (ECal), hadronic calorimeter (HCal) and muon system (MS)~\cite{Lee:2018pag}. The ID, ECal and HCal will be covered by strong magnetic field. For long-lived $H_5^{\pm\pm}$, we consider detecting the displaced objects within either ID or MS.
The analysis of displaced calorimeter deposits in ECal and HCal~\cite{Lee:2018pag} is not considered in this work as it requires dedicated calorimeter simulation and the design of a suitable trigger system, which are beyond the scope of this study.
Further, when the $H_5^{\pm\pm}$ travels outside the detector, it will be treated as heavy stable charged particle (HSCP). In particular, the trajectory of $H_5^{\pm\pm}$ will have smaller radius within the magnetic field\footnote{The trajectory of charged particles in magnetic field is given in Appendix~\ref{appendix:trajectory}.}, as it is doubly charged, which is included in the consideration of HSCP analysis. The far detectors are extensions of the ATLAS/CMS/LHCb/ALICE detectors. These far detectors can also probe the HSCP scenario when the LLP decays within the corresponding detector volume.

In current studies, we focus on the pair production of $H_5^{\pm\pm}$ via the process $p\,p\to H_5^{++}H_5^{--}$ at the LHC with $\sqrt{s}=14\,\rm TeV$ and $\mathcal{L}=300/3000\,{\rm fb}^{-1}$.
For a long-lived doubly charged particle in the final state, after proper event selections, the relevant background can be reduced to a negligible level. We thus assume that the signal region we considered is background-free.

\subsection{\texorpdfstring{$H_5^{\pm\pm}$}{H5pp} decays within the inner tracking detector}

\begin{table}
    \centering
\begin{tabular}{|c|c|c|}
\hline
 & \textbf{Electrons} & \textbf{Muons} \\
\hline
\hline
\textbf{Transverse momentum}  & $p_T > 42 \ \text{GeV}$ & $p_T > 40 \ \text{GeV}$ \\
\hline
\multirow{3}*{\textbf{Pseudorapidity} ($|\eta|$)}  & $|\eta| < 1.44$ &  \\
                               ~ & \textbf{or} &    $|\eta| < 2.4$   \\
                               ~ & $1.56 < |\eta| < 2.4$ & \\
\hline
\textbf{Isolation cone}  & $\Delta R < 0.3 \ $ & $\Delta R < 0.4 \ $ \\
\hline
\textbf{Isolation variable} &  $\frac{p^{\text{iso}}_T}{p_T} < 0.035$ ($|\eta| < 1.44$)  &   \\
\textbf{with} & \textbf{or} & $\frac{p^{\text{iso}}_T}{p_T} < 0.15$ ($|\eta| < 2.4$) \\
\textbf{its} $|\eta|$ \textbf{coverage} & $\frac{p^{\text{iso}}_T}{p_T} < 0.065 $ ($1.56<|\eta|<2.4$) & \\
\hline
\end{tabular}
\caption{Summary of the basic selections imposed on the candidate electrons and muons.}
\label{tab:mu_e_container}
\end{table}

We first focus on the long-lived $H_{5}^{\pm\pm}$ decays within the inner tracker system. Here, the leptonic decay modes of $H_{5}^{\pm\pm}$ are considered: $H_5^{\pm\pm} \rightarrow \ell^{\pm}\nu_{\ell}\ell^{\pm}\nu_{\ell}$ where $\ell = e, \mu$, as the signature of a displaced vertex with same-sign charged leptons can largely reduce possible background events.
The signal events are generated by {\tt MadGraph5}~\cite{Alwall:2011uj} and then passed to {\tt Pythia8}~\cite{Sjostrand:2014zea} for showering and hadronization. At the generator level, we applied the following basic cuts for the final state charged leptons: $p_T^{\ell}>10\,\rm GeV$, $|\eta_{\ell}|<2.5$ and $\Delta R_{\ell\ell}>0.4$.
To simulate the detector effects, we use the simplified-fast simulation (SFS) framework~\cite{Araz:2020lnp} embedded in {\tt MadAnalysis5}~\cite{Conte:2012fm}. This analysis utilized the CMS-EXO-16-022 template SFS card~\cite{Araz:2020lnp}, and the effect of the magnetic field on particle trajectory deflection within the SFS framework was determined for the transverse impact parameter, $d_{0,\ell}$.
The leptons (electron or muon) from the $H_5^{\pm\pm}$ decay should satisfy the basic event selections listed in~\autoref{tab:mu_e_container}~\cite{Araz:2021akd}, which require a energetic (minimum value for $p_T$) isolated charged leptons (requirement on $p_T^{\rm iso}$) that can be covered by the corresponding detector layers (range of $\eta$). Here $p_T^{\rm iso}$ is defined as the scalar sum of the transverse momentum of all reconstructed objects lying within a cone of a specified size $\Delta R < 0.3$ and is centered around the momentum of the leptons.
The recasting analysis~\cite{Araz:2021akd} in the Public Analysis Database (PAD)~\cite{Conte:2014zja,Dumont:2014tja,Conte:2018vmg,Araz:2019otb} are then adapted for our purpose for the long-lived $H_5^{\pm\pm}$ decaying within the ID.

\begin{figure}[!bp]
\centering{\includegraphics[width=0.6\textwidth]{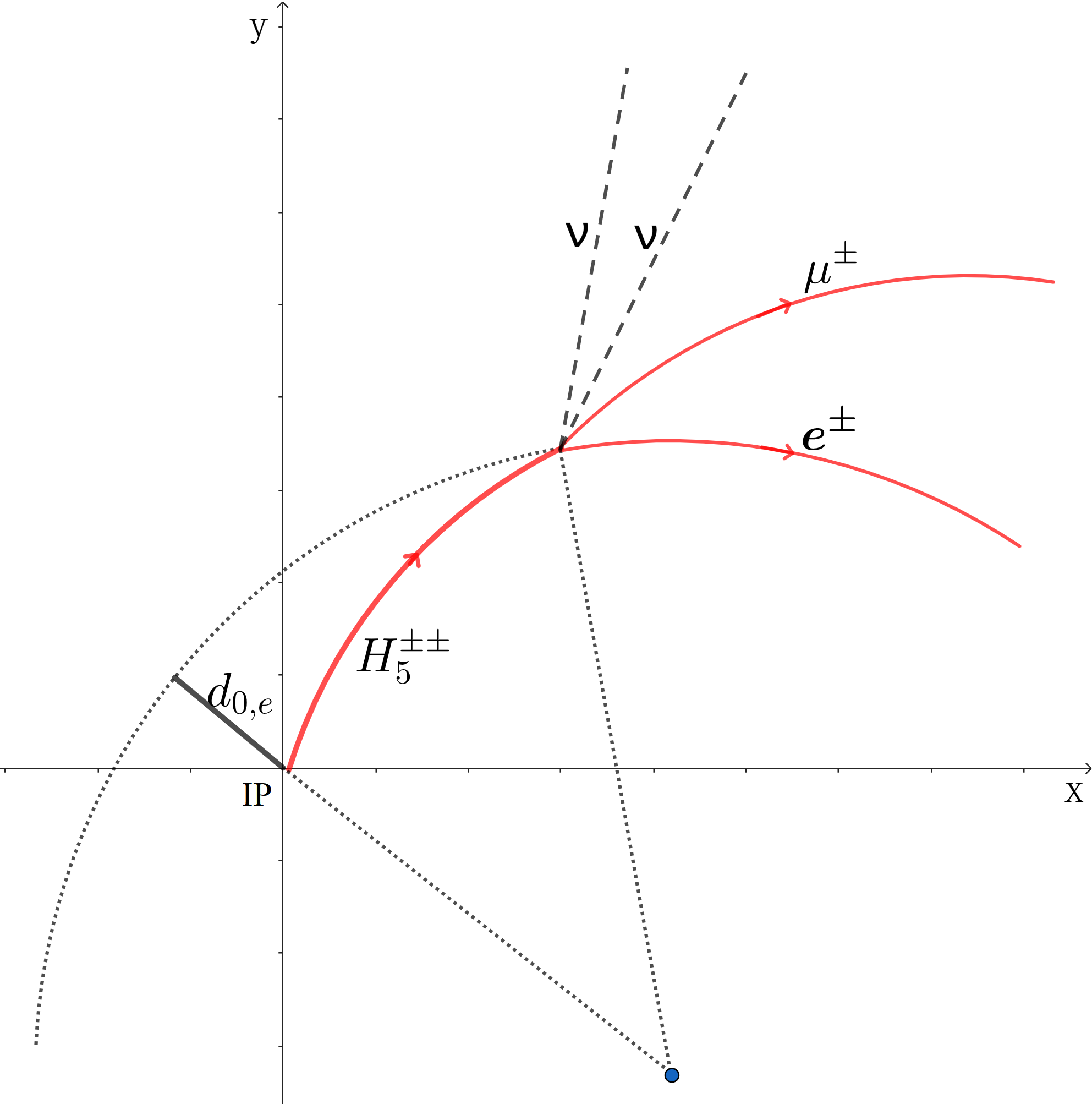}}
\caption{Definition of the transverse impact parameter $d_{0,\ell}$~\cite{Araz:2021akd} in the tracker system.}
\label{fig:displaced_leptons_d0}
\end{figure}

The following event selections are applied further in order to single out the signal events:
\begin{enumerate}
    \item At least one pair of charged leptons with the same charge,
    \item $0.2 < \lvert d_{0,\ell} \rvert < 100$  mm,
	\item ${\rho_{\ell}} < 10$ cm and ${z_\ell} < 30$ cm,
    \item $\Delta R_{\ell^{\pm}\ell^{\prime\pm}} < 1.5$,
	\item ${\slashed{E}_{T}} >$ $20$ GeV,
    \item Veto $0.7 < \Delta \phi^{miss}_{\ell^{\pm}\ell^{\prime\pm}} < 2.5$ with $\Delta \phi^{miss}_{\ell^{\pm}\ell^{\prime\pm}}$ $\equiv$ $\lvert \Delta \phi({\vec{p}_{T}^{ miss}},\vec{p}_T^{\ell^{\pm} \ell^{\prime\pm}})  \rvert$,
\end{enumerate}
where $\ell = e, \mu$ and $\rho_{\ell}$, ${z_\ell}$ are the transverse and longitudinal distance, with respect to the IP, of the lepton production vertex, defined as the intersection point of the tracks from the same-sign charged leptons.
$\Delta R_{\ell^{\pm}\ell^{\prime\pm}}$ is the distance in $\eta$-$\phi$ plane between two same-sign leptons,
${\slashed{E}_{T}}$ is the transverse missing energy, $\Delta \phi^{miss}_{\ell^{\pm}\ell^{\prime\pm}}$ represents the azimuthal angular separation between the missing transverse momentum (${\vec{p}_{T}^{\rm miss}}$) and the transverse momentum of a pair of same-sign charged leptons ($\vec{p}_T^{\ell^{\pm} \ell^{\prime\pm}}$),
$d_{0,\ell}$ is the transverse impact parameter for the corresponding charged lepton $\ell$ as shown in~\autoref{fig:displaced_leptons_d0} for $|d_{0,\ell}|$ in $H_5^{\pm\pm}\to \ell^\pm\nu_\ell \ell^\pm\nu_\ell$ case where both the $H_5^{\pm\pm}$ and the charged leptons will be bended by the strong magnetic field which is considered in the analysis according to Appendix~\ref{appendix:trajectory}.

\begin{table}[!bp]
\begin{center}
\begin{tabular}{|c|c|c|c|}
\hline
Cut flow in $\sigma$ [fb] & BP-1 & BP-2 & BP-3 \\
\hline Generator & $8.221$ & $3.927$ & $3.927$ \\
\hline Same-sign lepton pair $\geq1$ & $0.253$ & $0.383$ & $0.381$ \\
\hline $0.2< \lvert d_{0,\ell} \rvert <100$ mm & $0.251$ & $4.03\times10^{-2}$ & $0.356$ \\
\hline $\rho_\ell<10$ cm, $z_\ell<30$ cm & $2.40\times10^{-2}$ & $3.82\times10^{-2}$ & $3.47\times10^{-2}$ \\
\hline $\Delta R_{\ell^\pm\ell^{\prime\pm}} < 1.5$ & $2.31\times10^{-2}$ & $3.70\times10^{-2}$ & $3.31\times10^{-2}$ \\
\hline $\slashed{E}_T > 20$ GeV & $2.09\times10^{-2}$ & $3.34\times10^{-2}$ & $3.09\times10^{-2}$ \\
\hline $\Delta\phi$ veto & $1.81\times10^{-2}$ & $3.12\times10^{-2}$ & $2.92\times10^{-2}$ \\
\hline
\end{tabular}
\caption{The cut-flow table for $H_5^{\pm\pm}$ leptonic decay within ID for the three benchmark points. The details of each selection are listed in the main text.}
\label{tab:CF_displaced_leptons_70}
\end{center}
\end{table}
\begin{figure*}[ht!]
    \centering
    \includegraphics[width=0.48\textwidth]{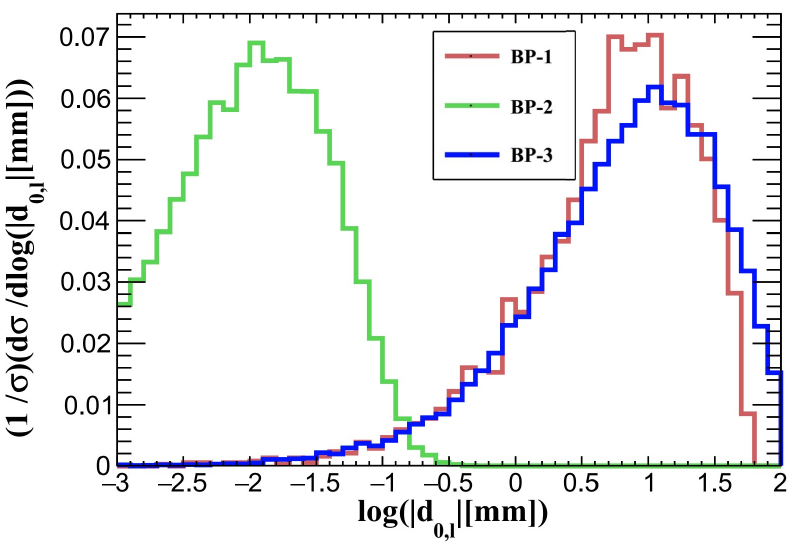}
    \includegraphics[width=0.48\textwidth]{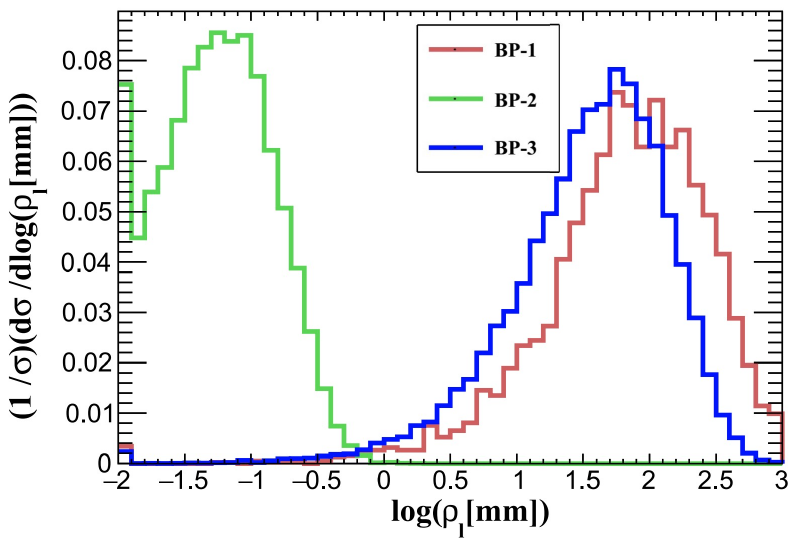}
    \includegraphics[width=0.48\textwidth]{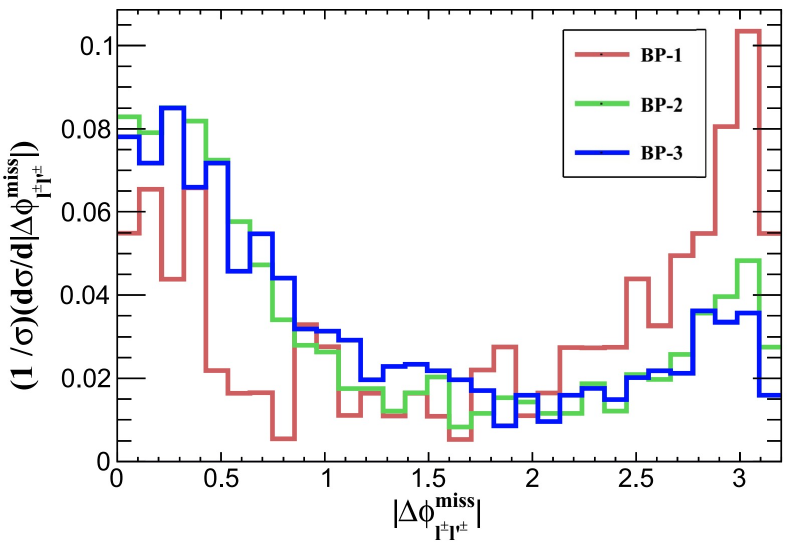}
    \includegraphics[width=0.48\textwidth]{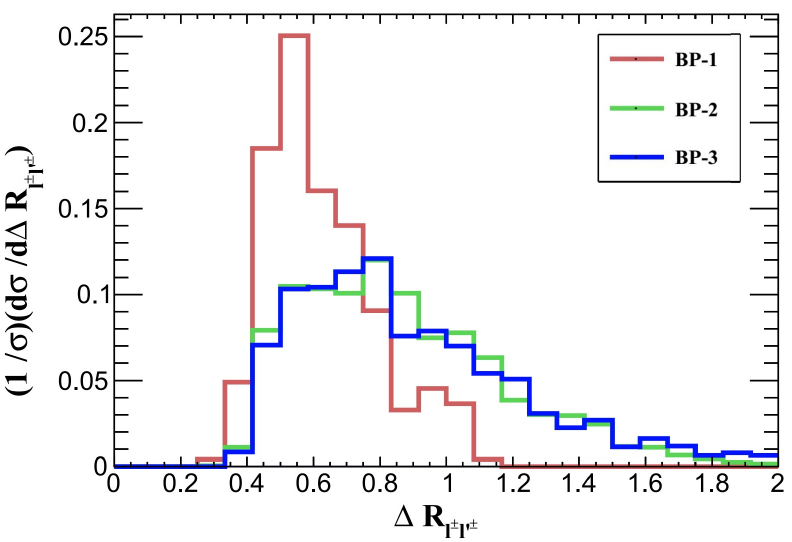}
	\caption{$d_{0,\ell}$, $\rho_\ell$, $|\Delta\phi_{\ell^\pm\ell^\prime\pm}^{\rm miss}|$ and $\Delta R_{\ell^{\pm}\ell^{\prime\pm}}$ distributions for $H_5^{\pm\pm}$ leptonic decay within ID for three benchmarks: BP-1 (red) $m_5=70\,{\rm GeV}, s_H = 5\times10^{-5}$; BP-2 (green) $m_5=130\,{\rm GeV}, s_H = 5\times10^{-5}$ and BP-3 (blue) $m_5=130\,{\rm GeV}, s_H = 8\times10^{-7}$.
     }
	\label{fig:displaced leptons_130}
\end{figure*}

Pair production of leptonic decay $H_5^{\pm\pm}$ provides at least one pair of same-sign charged leptons. In order to suppress the background from multi-gauge boson productions, we observed that the decay products of $H_5^{\pm\pm}$ tend to be collinear, resulting in a smaller separation between the two same-sign charged leptons. On the other hand,  pair of $H_5^{\pm\pm}$ are produced back-to-back in the transverse plane, the transverse momentum of the pair of two same-sign charged leptons, $\vec{p}_T^{\ell^\pm\ell^{\prime\pm}}$, is close to either the same or opposite direction of $\vec{p}_T^{\rm miss}$. Therefore, we require $\Delta R_{\ell^\pm\ell^{\prime\pm}}<1.5$ for the pair of same-sign charged leptons and veto $0.7<\Delta\phi_{\ell^\pm\ell^{\prime\pm}}^{\rm miss}<2.5$ to suppress the multi-gauge boson backgrounds. Further, the long-lived nature of $H_5^{\pm\pm}$ and the requirement of the decay within ID provide that $\rho_\ell<10\,\rm cm$ and $z_\ell<30\,\rm cm$ which are adapted from~\cite{Araz:2021akd}, and also $0.2<|d_{0,\ell}|<100\,\rm mm$. The $d_{0,\ell}$, $\rho_\ell$, $\Delta R_{\ell^\pm\ell'^\pm}$ and $\Delta\phi_{\ell^\pm\ell'^\pm}^{\rm miss}$ distributions are shown in~\autoref{fig:displaced leptons_130} for three benchmarks:
\begin{itemize}
    \item BP-1: $m_5=70$ GeV, $s_H=5\times10^{-5}$,
    \item BP-2: $m_5=130$ GeV, $s_H=5\times10^{-5}$,
    \item BP-3: $m_5=130$ GeV, $s_H=8\times10^{-7}$.
\end{itemize}
Note that, the $s_H$ for BP-3 is chosen such that it has roughly the same $c\tau$ as BP-1. BP-2 has smaller $c\tau$ than BP-1 and BP-3. Hence, BP-2 has much smaller $d_{0,\ell}$ and $\rho_{\ell}$ than BP-1 and BP-3 as can be seen from~\autoref{fig:displaced leptons_130}. On the other hand, other kinematic distributions are not sensitive to the $c\tau$ and depends only on the mass of the $H_5^{\pm\pm}$.
The cut-flow table for the cross section of above three benchmark points are shown in~\autoref{tab:CF_displaced_leptons_70}.
With these cuts, especially the requirement of a same-sign lepton pair, the range for $\rho_\ell$ and $d_{0,\ell}$, the background is assumed to be negligible.

The number of signal event is then obtained as $N_S = \epsilon\times\sigma\times\mathcal{L}$ where $\epsilon$ is the selection efficiency calculated for different $m_5$ and $s_H$, $\sigma$ is the signal cross section without any cut. In this study, we assume $\mathcal{L} = 300/3000\,{\rm fb}^{-1}$. For the background-free case, $N_S = 3$ defines the 95\% confidence level (CL.) exclusion limit.
We then extend the same search strategies over a wide range of $m_5$ values and the 95\% exclusion regions in $m_5$-$c\tau$ plane as well as $m_5$-$s_H$ plane are shown in the blue area of~\autoref{fig:LLP_Exclusion}.
One can see the searches from inner tracker can cover the $c\tau$ from $10^{-4}\,\rm m$ to about $1\,\rm m$. The coverage becomes weak at lower mass, as the efficiency drops due to a fixed basic cuts on the final states.

\subsection{\texorpdfstring{$H_5^{\pm\pm}$}{H5pp} decays within the muon system}

If the lifetime of $H_5^{\pm\pm}$ is much longer, it may travel through the tracker and calorimeters and decay within the muon system. In particular, in the following analysis, we consider the case where the decay happens in the CMS endcap muon detectors (EMDs) where the cathode strip chambers (CSCs) are installed. This analysis uses the CMS-CSCCluster template card~\cite{mitridate2023energetic} in Delphes to simulate the efficiency of detector reconstruction clusters and the effect of the magnetic field on the charged trajectories of LLPs within the detector.
The detailed configurations of the detectors can be found in~\cite{CMS:1997iti,CMS:2008xjf,CMS:2021juv}. The analysis follows the CMS searches for LLPs decaying in the EMDs~\cite{CMS:2021juv} and the recasting analysis which tabulates the efficiency of CSCs of the CMS experiments~\cite{Cottin:2022nwp}~\footnote{The tabulated efficiency of CSC is available in relevant modules in {\tt Delphes}, see \url{https://github.com/delphes/delphes/pull/103}. }.
Using the CSCs for LLP searches can efficiently reduce the background to a sufficiently low level thanks to a large amount of absorber material in front of the EMD acting as a shield. Further, the CSCs as sampling calorimeter are more sensitive to the LLP energy rather than its mass. It hence provides equally sensitivity to all LLP masses considered here.

We consider all possible decay channels of $H_5^{\pm\pm}$ in this case: leptonic ($H_5^{\pm\pm}\to \ell^{\pm}\nu_{\ell}\ell^{\pm}\nu_{\ell}$), hadronic ($H_5^{\pm\pm}\to jjjj$) and semi-leptonic ($H_5^{\pm\pm}\to \ell^{\pm}\nu_{\ell}jj$). Note that the muon can penetrate the shield easily, for the above channels, we thus require the signal signature with at least a pair of same charged muons or a {\it CSC cluster} from one $H_5^{\pm\pm}$ in order to suppress relevant background events.
At the generator level, we required $p_T^{\ell}>10\,\rm GeV$, $|\eta_{\ell}|<2.5$ and $\Delta R_{\ell\ell}>0.4$ for leptons, and $p_T^j>20\,\rm GeV$, $|\eta_j|<5.0$, $\Delta R_{jj}>0.4$ and $\Delta R_{j\ell}>0.4$ for jets.
In this case, $H_5^{\pm\pm}$ is assumed to decay into jets or leptons within the muon system, and its decay products will leave signals in the CSC detectors.
The response of the CSC detector is simulated with relevant modules in {\it delphes} using the tabulated efficiency extracted from the CMS results~\cite{CMS:2021juv}.
After reconstruction, the final states contain muons and/or {\it CSC cluster} which is the group of the signals from different layers/components of {\it CSCs}. The final state muon is reconstructed in the usual way, but is required to have $1.5<|\eta|<2.4$ within the coverage of CSC detectors. The muon should further satisfy the isolation requirements: $\frac{\sum_{i} p_T^{\text{particle}_i}}{p_T^{\mu}} \leq 0.25$, where the summation is over all particles inside the isolation cone $\Delta R = \sqrt{(\Delta \eta)^2 + (\Delta \phi)^2} \leq 0.5$ with $p_T^{\text{particle}} > 0.5$ GeV, and $p_T^{\mu}$ is the transverse momentum of the muon.

The reconstructed {\it CSC cluster} is required to satisfy the following requirements:
\begin{itemize}
    \item Containing at least 50 hits in the CSC detectors,
    \item CSC cluster time between -5 ns and 12.5 ns to reject clusters produced by pileup,
    \item All clusters within a distance of $\Delta R = \sqrt{(\Delta \eta)^{2} + (\Delta \phi)^{2}} < 0.4$ from a muon are removed to ensure that the {\it CSC cluster} is not associated with muons having $p_T \geq 10$ GeV, and to prevent the cluster from being generated by muon bremsstrahlung~\cite{liu2024revealing}.
\end{itemize}
In the current case, the {\it CSC cluster} is induced either by the electron or the jet from the $H_5^{\pm\pm}$ decay. However, any such decay products from the same $H_5^{\pm\pm}$ (and thus traveling in similar directions) will result in at most one {\it CSC cluster}, even if both hit the same groups of {\it CSC} detectors. For {\it CSC cluster} induced by various combinations of the electron and/or jet decay products from $H_5^{\pm\pm}$, the reconstruction efficiencies are listed in~\autoref{tab:cluster_Eff} for two mass benchmarks, $m_5 = 70$ GeV and $130$ GeV. It is evident that with higher energy from $H_5^{\pm\pm}$ due to a heavier mass, the efficiency increases.

\begin{table}[!tbp]
	\begin{center}
		\begin{tabular}{|c|c|c|c|c|c|}
			\hline Type & $e^{\pm}$ & $e^{\pm}e^{\pm}$ & $j j$ & $j j j j$ & $j j e^{\pm}$ \\
			\hline Efficiency for $m_5=70\,\rm GeV$ & 7.6\% & 8.2\% & 2.9\% & 3.2\% & 8.0\% \\
			\hline Efficiency for $m_5=130\,\rm GeV$ & 10.7\% & 11.7\% & 4.0\% & 5.1\% & 11.2\% \\
			\hline
		\end{tabular}
		\caption{The reconstruction efficiency of {\it CSC cluster} induced from different decay products of $H_5^{\pm\pm}$ for two mass benchmarks with $m_5=70\,\rm GeV$ and $m_5=130\,\rm GeV$.}
		\label{tab:cluster_Eff}
	\end{center}
\end{table}

The analysis is classified into four channels according to the number of {\it CSC cluster} $N_C$ and the number of same-sign muon pair $N_{\mu}^{SS}$:
\begin{enumerate}
    \item $N_C = 2, N_\mu^{SS}=0$, where both $H_5^{\pm\pm}$ decay into electrons or jets and are all reconstructed as {\it CSC cluster}.
    \item $N_C = 1, N_\mu^{SS}=0$, where only one of $H_5^{\pm\pm}$s is reconstructed as {\it CSC cluster} and the other one escapes from the detector.
    \item $N_C = 1, N_\mu^{SS}=1$, where one of $H_5^{\pm\pm}$s is reconstructed as {\it CSC cluster} and the other decays into a pair of muons.
    \item $N_C = 0, N_\mu^{SS}=1$, where one $H_5^{\pm\pm}$ is reconstructed through its muonic decay channel, and the other one escapes from the detector.
    \item $N_C = 0, N_\mu^{SS}=2$ (one positive and one negative muon pair), where both $H_5^{\pm\pm}$ decay into a pair of muons.
\end{enumerate}
The signal cross sections of these five channels for two benchmark points $m_5 = 70$ GeV with $s_H = 3.2\times10^{-5}$ and $m_5 = 130$ GeV with $s_H = 3.8\times10^{-7}$ are listed in~\autoref{tab:muon_chamber_cs}, where the requirement that $H_5^{\pm\pm}$ should decay within the CSC detector is also included. As mentioned above, we do not include the case where one $H_5^{\pm\pm}$ only produces one muon together with other particles. However, we have also checked the event rate from such signature, which only provides several percent increase in the total rate. We then extend the same search strategies over a wide range of $m_5$ values and the 95\% exclusion regions in $m_5$-$c\tau$ plane as well as $m_5$-$s_H$ plane are shown in the pink area of~\autoref{fig:LLP_Exclusion}. Notably, the searches with CSC detector of the muon system can cover the $c\tau$ of $H_5^{\pm\pm}$ from below meter to about 10-100 meters almost independent of the $H_5^{\pm\pm}$ mass.

\begin{table}[!tbp]
    \begin{center}
        \begin{tabular}{|c|c|c|}
            \hline  \multirow[c]{2}{*}{ $\sigma$ [fb]} & ~$m_5=70$ GeV~ & ~$m_5=130$ GeV~\\
            & $s_H = 3.2\times 10^{-5}$ & $s_H=3.8\times10^{-7}$\\
            \hline ~$N_C=2,N_\mu^{SS}=0$~ & $1.80\times10^{-3} $ & $6.44\times10^{-3} $ \\
            \hline $N_C=1,N_\mu^{SS}=0$ & $2.71\times10^{-2} $ & $1.00\times10^{-1} $ \\
            \hline $N_C=1,N_\mu^{SS}=1$ & $6.36\times10^{-4} $ & $7.48\times10^{-3} $ \\
            \hline $N_C=0,N_\mu^{SS}=1$ & $7.99\times10^{-4} $ & $9.56\times10^{-2} $ \\
            \hline $N_C=0,N_\mu^{SS}=2$ & $2.32\times10^{-4} $ & $1.53\times10^{-2} $ \\
            \hline Total & $3.10\times10^{-2} $ & $2.24\times10^{-1} $ \\
            \hline
        \end{tabular}
        \caption{
            The cross section for $H_5^{\pm\pm}$ decaying in CSC detector for four different channels of two benchmark points: $m_5 = 70$ GeV with $s_H = 3.2\times10^{-5}$ and $m_5 = 130$ GeV with $s_H = 3.8\times10^{-7}$.
        }
        \label{tab:muon_chamber_cs}
    \end{center}
\end{table}

\subsection{\texorpdfstring{$H_5^{\pm\pm}$}{H5pp} as a heavy stable charged particle}
\label{sec:HSCP}

If the $H_5^{\pm\pm}$ has even longer lifetime, it can travel through the whole detector before its decay, there is no explicit signal except the charged tracks in the detector, rendering it as heavy stable charged particle (HSCP). Without a dedicated search strategy, there is a risk of misidentification or complete oversight of HSCPs, as the particle identification algorithm in hadron collider experiments is often tailored to signature characteristics of SM particles. Both ATLAS and CMS have performed detailed analysis about the HSCP with full detector simulations which put strong constraints on the production cross section of such particles~\cite{chatrchyan2013searches,cms2015constraints}.

However, in order to apply the relevant results on GM model, one needs dedicated simulation of the detector response for the doubly charged $H_5^{\pm\pm}$ which is beyond the scope of current work. The CMS Collaboration proposed a ``fast technique''~\cite{CMS:2015lsu} to simulate the response of the detector to HSCPs which is embedded in an efficiency table as function of $p_T$, $\eta$ and $\beta$ of the charged tracks. Nevertheless, this efficiency table is currently provided only for singly charged HSCP. For doubly charged HSCP, the energy deposition will change dramatically which alters significantly the detector efficiency. On the other hand, doubly charged HSCP, with the same momentum, will have smaller tracking radius compared to singly charged HSCP. Combining all of these factors, it is not reliable to use this efficiency table for our case. Hence, for the current case, we do not estimate the prospects of detecting $H_5^{\pm\pm}$ as a HSCP , as this would require a detailed simulation of the detector response. Instead, we simply apply the previous constraints on the production cross section from the 7/8 TeV analysis~\cite{CMS:2013czn}. The analysis performed in~\cite{CMS:2013czn} for multiple charged HSCPs utilizes the energy deposition information in the detector and the time-of-flight (TOF) measurement from the muon system. The multiply charged particles will produce greater ionization in the detector compared to singly charged particles. The $\beta$ inferred from the TOF measurement for a relatively heavy particle will be smaller than that of SM particles. In order to apply the constraint in our case, for each mass of $H_5^{\pm\pm}$ and its $c\tau$, we calculate the cross section of $H_5^{\pm\pm}$ production that can flight out of the detector. With varying $\eta$, the minimum travel distance required to traverse the entire detector will differ slightly. Here, we follow the conservative estimation of such threshold from~\cite{CMS:2015lsu} requiring
\begin{align}
    L \geq \begin{cases}
        9.0\,{\rm m}, & 0.0\leq|\eta|\leq 0.8,\\
        10.0\,{\rm m}, & 0.8\leq|\eta|\leq 1.1,\\
        11.0\,{\rm m}, & 1.1\leq|\eta|.
    \end{cases}
\end{align}
The traveling distance can be estimated statistically event by event according to the momentum of $H_5^{\pm\pm}$ and its $c\tau$. In the lab frame, the probability of $H_5^{\pm\pm}$ can travel longer than some threshold $L$ is given by $\exp\left(-L/\gamma\beta\tau\right)$. The production cross section $H_5^{\pm\pm}$ is scaled by such probability and we ignore the case where $H_5^{\pm\pm}$ decay before travelling above the threshold distance which is already covered by the analysis in the above sections. The cross section is then compared with the upper limit provided by the CMS~\cite{CMS:2013czn} with $\sqrt{s}=7/8\,\rm TeV$. The results are shown in yellow area ($\sqrt{s}=7\,\rm TeV$) and orange area ($\sqrt{s}=8\,\rm TeV$) of~\autoref{fig:LLP_Exclusion} with coverage only above $m_5>100\,\rm GeV$. The region above (below) corresponding $c\tau$ ($s_H$) can all be excluded. The exclusion line in $c\tau$ is well aligned with the size of the detector. Therefore, we expect that even with higher energy and higher luminosity, the exclusion will not change too much.

\subsection{\texorpdfstring{$H_5^{\pm\pm}$}{H5pp} decays in far detectors}

\begin{figure}
\centering
\includegraphics[width=\textwidth]{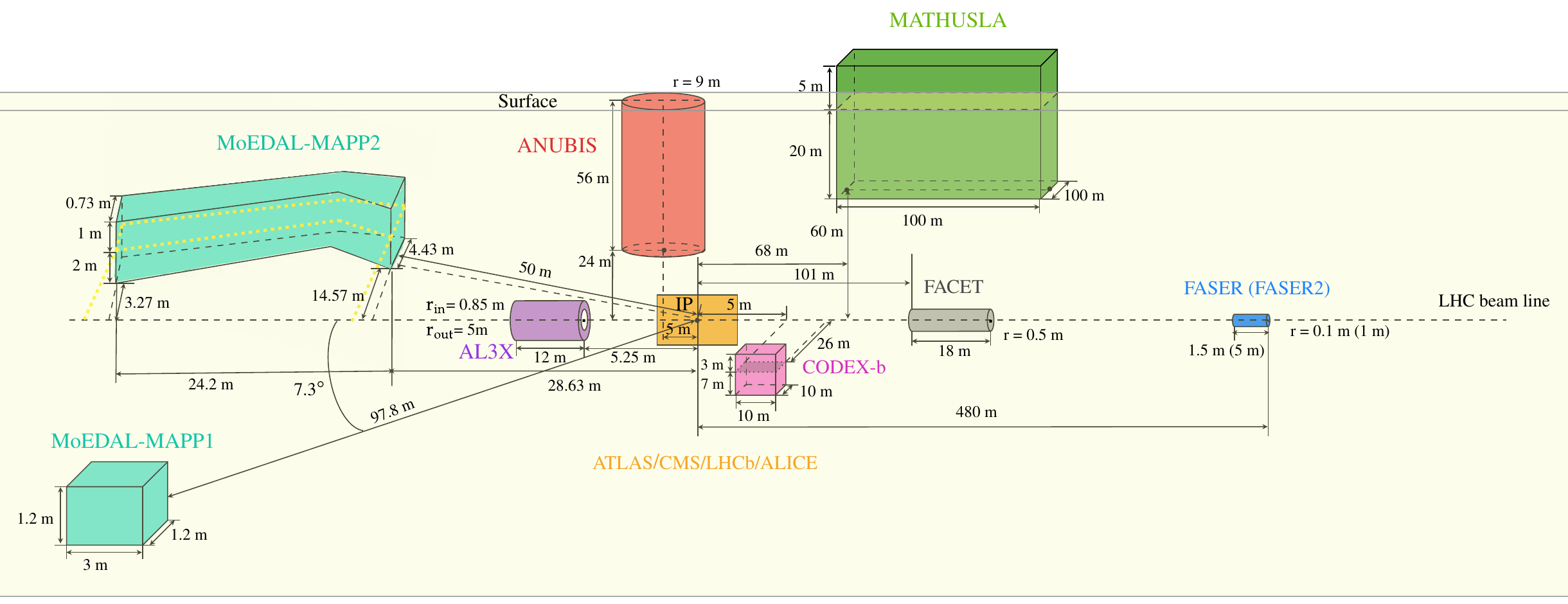}
\caption{The positions of far detectors relative to ATLAS/CMS/LHCb/ALICE detectors including ANUBIS, MATHUSLA, FACET, FASER, COEDX-b, MoEDAL-MAPP and AL3X. To better illustrate the position and size of the detectors, the picture is not drawn to scale, however, the marked values reflect the actual dimensions.}
\label{fig:far_detectors_picture}
\end{figure}

Similar to the case discussed in~\autoref{sec:HSCP}, $H_5^{\pm\pm}$ can traverse the entire detectors with a sufficiently long lifetime. In addition to treating it as a HSCP, various detectors located farther from the interaction point (IP) can be used to identify signals from such LLPs. These detectors include ANUBIS~\cite{Bauer:2019vqk}, MATHUSLA~\cite{MATHUSLA:2020uve}, FACET~\cite{Cerci:2021nlb}, FASER/FASER2~\cite{FASER:2018eoc}, CODEX-b~\cite{Gligorov:2017nwh}, MoEDAL-MAPP~\cite{Pinfold:2019nqj} and AL3X~\cite{Gligorov:2018vkc}. The locations of these detectors relative to the ATLAS/CMS/LHCb/ALICE detectors are shown in~\autoref{fig:far_detectors_picture}, where we indicate the distance to the IP and the geometric size of each detector.

\begin{itemize}
\item MAssive Timing Hodoscope for Ultra-Stable neutraL pArticles, MATHUSLA~\cite{MATHUSLA:2020uve,MATHUSLA:2018bqv} is a box-shaped detector with dimensions of 100 m $\times$ 100 m $\times$ 25 m for detecting LLPs. It is located about 70 m away from CMS IP, with approximately 20 m of decay volume below surface and 5 m of decay volume above surface.
\item AN Underground Belayed In-Shaft, ANUBIS~\cite{Bauer:2019vqk,Shah:2024fpl,Dreiner:2020qbi} is a proposed cylindrical detector that uses the ATLAS installation shafts with a diameter of $18$ m and a length of $56$ m for experimentation. The detector is placed 24 m above the ATLAS IP with 5 m offset.
\item Forward-Aperture CMS ExTension, FACET~\cite{Cerci:2021nlb} is a cylindrical detector with $0.5$ m radius and 18 m length located $119$ m away from the CMS IP along the longitudinal direction.
\item The Forward Search Experiment, FASER~\cite{FASER:2018eoc,Domingo:2023dew} is designed to search light particles with extremely weak interactions. The detector is located 480 m away from the ATLAS IP along the longitudinal direction. FASER detector, which is already installed, is a cylindrical detector with 0.1 m radius and 1.5 m length. The future upgrade FASER2 will extend its radius to 1 m and length to 5 m.
\item COmpact Detector for EXotics at LHCb, CODEX-b~\cite{Gligorov:2017nwh,Aielli:2019ivi} is a $10\times10\times10\,{\rm m}^3$ cube detector located $25\,\rm m$ away from the LHCb IP which covers the range of $x\in[26,36]\,\rm m$, $y\in[-7,3]\,\rm m$ and $z\in[5,15]\,\rm m$, where the z-axis is along the beamline at the IP, the y-axis is perpendicular to the ground.
\item MoEDAL’s Apparatus for Penetrating Particles, MAPP~\cite{Pinfold:2019nqj,Deppisch:2023sga,Kalliokoski:2023cgw} is a detector designed for LLPs in the MoEDAL~\cite{Felea:2020cvf,Acharya:2020uwc,Hirsch:2021wge} experiment frame. Phase 1 of MAPP(MAPP1) with $30\, {\rm fb}^{-1}$ is located in the UA83 tunnel about 97.8 m from the LHCb IP, with a $7.3^\circ$ angle relative to the beam direction. Its dimension is 1.2m$\times$1.2m$\times$3m. Phase 2 of MAPP(MAPP2) with $300\,{\rm fb}^{-1}$ is a detector planned to be built in the UGC1 tube approximately 50m from the LHCb IP. It is composed of two polyhedra.
\item A Laboratory for Long-Lived eXotics, AL3X~\cite{Gligorov:2018vkc,Dercks:2018wum} is a cylindrical detector located in the LHC-ALICE/L3 cave, $5.25\,\rm m$ from the collision site. It has a length of $12\,\rm m$ and an inner and outer radius of 0.85 and $5\,\rm m$, respectively.
\end{itemize}

For these far detectors, we assume that whenever the LLP decays within the detector volume, it can be detected. To evaluate the sensitivity of the far detectors, the {\tt Display Decay Counter} (DDC)~\cite{Domingo:2023dew} is utilized to estimate the probability for a given LLP decaying within the corresponding detectors. In particular, along its trajectory, the probability that the particle decays traveling a distance between $\ell_1$ and $\ell_2$ in the lab frame is given by
\begin{align}
\mathcal{P}(\ell_{1},\ell_{2})=\exp\left[-\frac{\ell_{1}}{\gamma\beta\tau}\right]-\exp\left[-\frac{\ell_{2}}{\gamma\beta\tau}\right]
\end{align}
where $\beta$ is the velocity of the particle, $\gamma=(1-\beta^2)^{-1/2}$, $\tau$ is the proper lifetime of the particle. For a given event containing the LLP, $\ell_1$ and $\ell_2$ can be determined by the direction of the LLP and the boundaries of the corresponding detector.

\begin{figure}[!tbp]
	\centering{\includegraphics[width=0.49\textwidth]{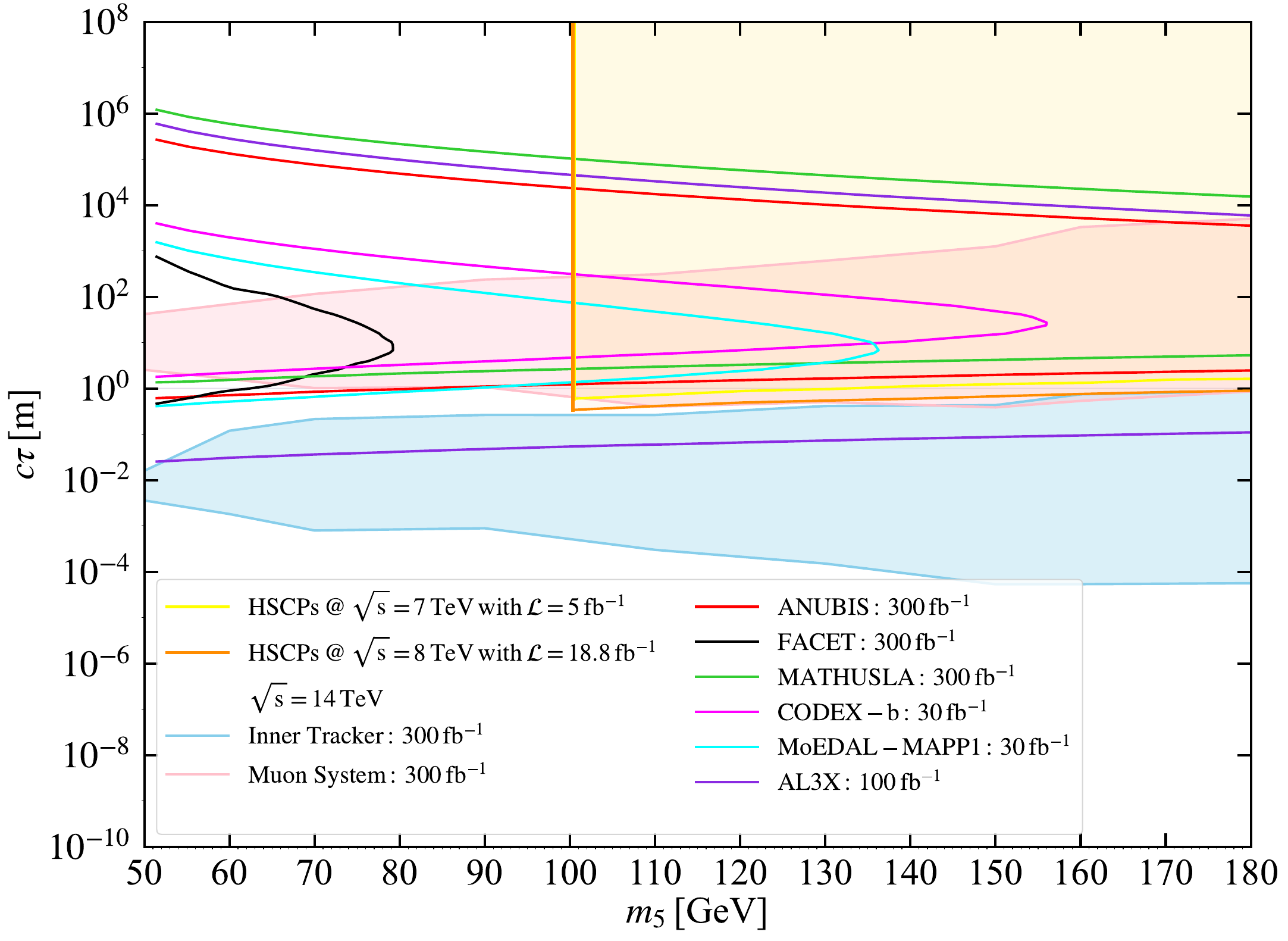}}
    \centering{\includegraphics[width=0.49\textwidth]{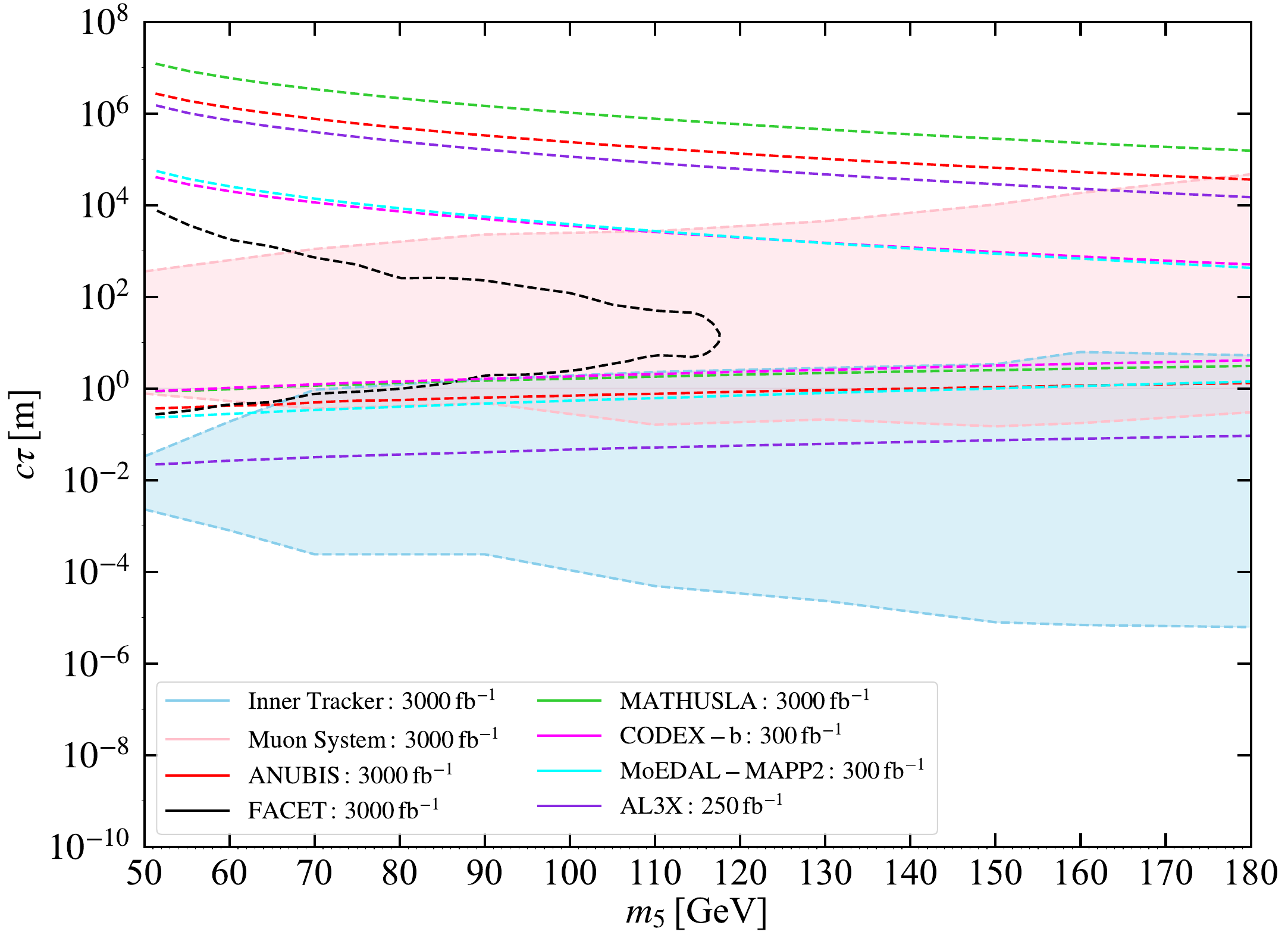}}
	\centering{\includegraphics[width=0.49\textwidth]{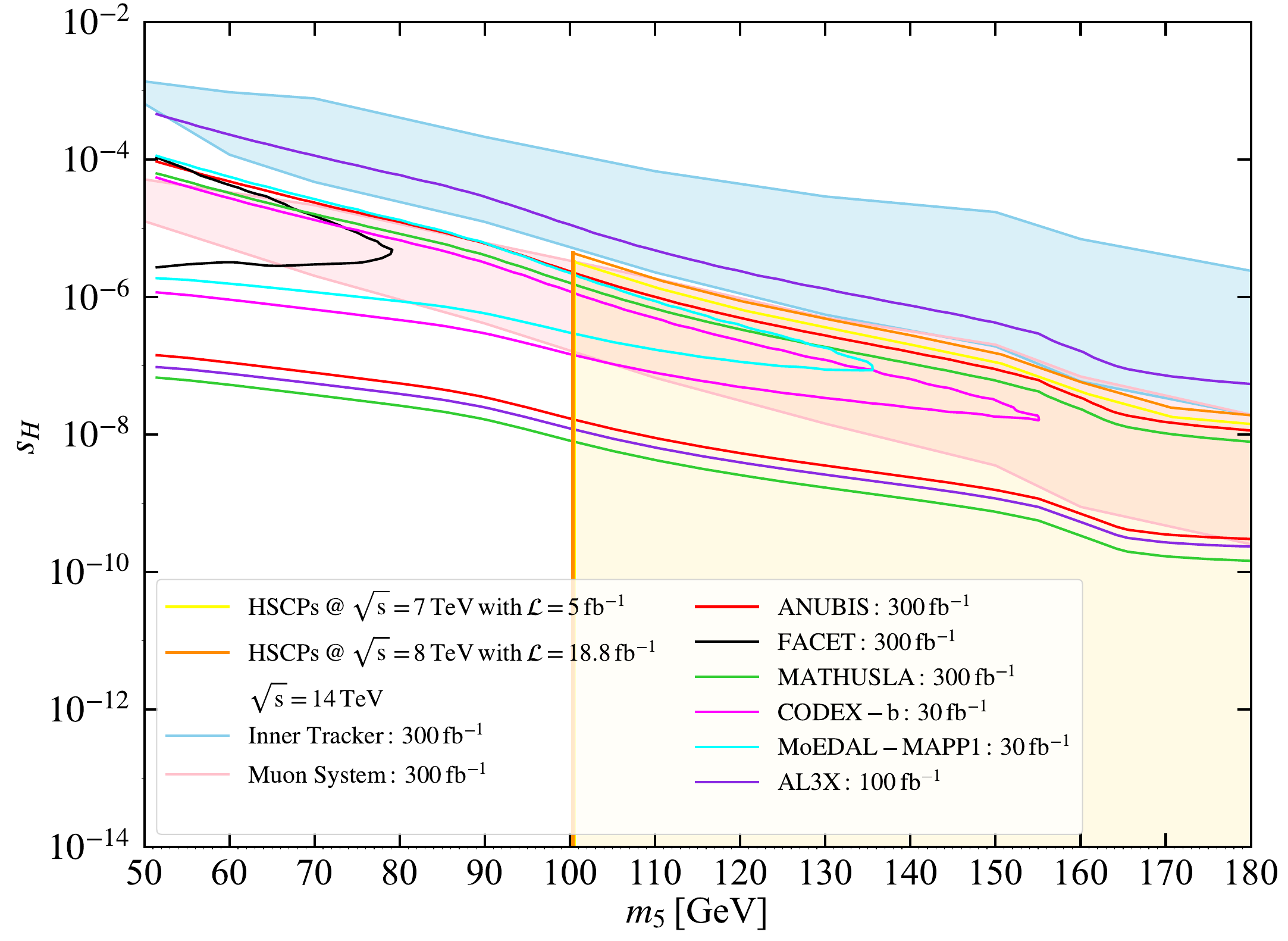}}
    \centering{\includegraphics[width=0.49\textwidth]{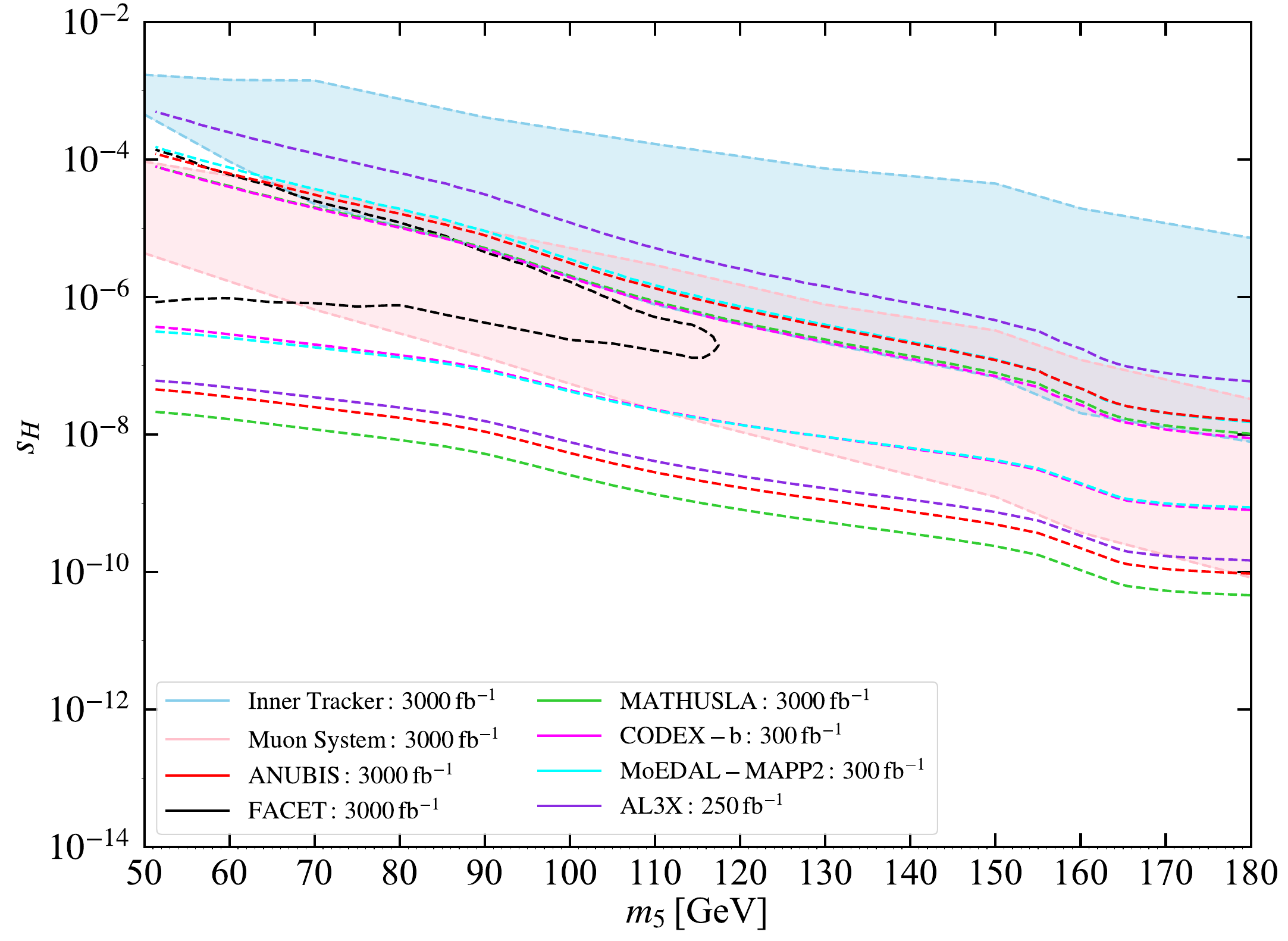}}
	\caption{The $95\%$ C.L. exclusion region for the search of long-lived $H_{5}^{\pm\pm}$ are shown in the $m_5$-$c\tau$ plane at the LHC (upper-left) and the HL-LHC (upper-right), as well as in the $m_5$-$s_H$ plane at the LHC (lower-left) and the HL-LHC (lower-right). Additionally, the analysis of HSCP for $\sqrt{s}=7$ TeV with $\mathcal{L}=5\ \text{fb}^{-1}$ and $\sqrt{s}=8$ TeV with $\mathcal{L}=18.8\ \text{fb}^{-1}$ are also shown in the left panels.}
	\label{fig:LLP_Exclusion}
\end{figure}

Considering that the signal from the IP has strong directional information and that the thick rock coverage in between will block all SM particles, the analysis can be considered to be background-free. The coverages for various far detectors in both $m_5$-$c\tau$ and $m_5$-$s_H$ planes are shown in~\autoref{fig:LLP_Exclusion}. The region within the corresponding lines is excluded at 95\% CL. with $\mathcal{L}=300/3000\,{\rm fb}^{-1}$ for ANUBIS, MATHUSLA, FACET and FASER/FASER2, $\mathcal{L}=30/300\,{\rm fb}^{-1}$ for CODEX-b and MoEDAL-MAPP, $\mathcal{L}=100/250\,{\rm fb}^{-1}$ for AL3X, respectively. Several comments are in order. The AL3X, FACET and FASER (FASER2), which are all detectors in forward region. However, AL3X is closer to the IP than FACET and FASER (FASER2) and has larger size and thus has better sensitivity. The very forward FACET and FASER (FASER2) have weaker sensitivity. They are in general more sensitive to the case where the BSM particles are produced forwardly which is not the case for $H_5^{\pm\pm}$ in current analysis. The MoEDAL-MAPP1 and CODEX-b are two detectors located near the LHCb. For MoEDAL-MAPP1, due to its small size and long distance from IP, it only covers the mass range of $m_5\in(50, 135)\,\rm GeV$. However, for CODEX-b, its size and distance from IP are moderate among all these detectors, the coverage in mass extends to $m_5\in (50, 160)\,\rm GeV$. The MATHUSLA and ANUBIS detectors have much better sensitivity as they are large in size and close to the IP. Besides, the ANUBIS is closer to the IP than that of MATHUSLA, hence the exclusion region is shifted towards lower (larger) $c\tau$ ($s_H$) compared with that of MATHUSLA. Furthermore, the boundaries of the exclusion region follow well with the contours of $c\tau$ of $H_5^{\pm\pm}$ shown in~\autoref{fig:ctau_m5_sH} with an extension at low mass region due to its much larger cross section.

\section{Conclusion}
\label{Sec:Conclusion}
The GM model extends SM model with real and complex triplets. The introduction of the complex triplet results in the doubly charged scalar $H_{5}^{\pm\pm}$ after EWSB. Currently the searches for the $H_5^{\pm\pm}$ are mainly focusing on high mass regions. However, the $c\tau$ of $H_5^{\pm\pm}$ increases with the decrease of both $m_5$ and $s_H$. Hence in the low-mass and low-$s_H$ region, the $H_5^{\pm\pm}$ can become long-lived, where the current searches at colliders mainly depending on the prompt final states signal are no longer applicable. In this work, we thus focus on such region where the doubly charged scalar is long-lived, and consider several different channels to cover the parameter region.

The searches can be categorized according to the lifetime of $H_5^{\pm\pm}$. When the lifetime is short, it may dominantly decay within the inner tracker, leaving displaced vertices for its charged decay products. For simplicity, we consider only the same-sign leptonic channel. The analysis for hadronic final states will be similar. Such channel can cover $c\tau$ from about $\mathcal{O}(10^{-4})$ meter to $\mathcal{O}(1)$ meter. Currently analysis assumes fixed selection rules for the charged leptons. Hence, the sensitivity is a bit weaker at lower mass where the leptons are also softer.

When the lifetime gets longer, $H_5^{\pm\pm}$ may travel through the innner tracker and the calorimeters and decay inside the muon system. For this case, we considered the CSC detector of the muon system. All decay products except muons, neutrinos will be reconstructed as the {\it CSC cluster} while the muon will be isolated as a charged track in the CSC detector. We consider all cases with different number of reconstructed {\it CSC cluster} and pair of same-sign muons. Combining all these cases, the searches can cover $c\tau$ from $\mathcal{O}(1)$ meter to $\mathcal{O}(100)$ meters.

If the lifetime of $H_5^{\pm\pm}$ is even longer, it may transverse the whole detector before it decays leaving charged track and be called HSCP. Such analysis relies heavily on the simulation of detector response to the charged particles transversing the detector which is beyond the scope of current work. However, by recasting the 7/8 TeV CMS results, we found that the searches for HSCP are powerful enough to cover the entire parameter space for a given mass, where the $c\tau$ is larger than several meters. Translating into the parameter space of GM model, it can cover, for given mass, the scenario that extremely closes to the alignment limit in the GM model, which shows an important complementarity to the other searches at the LHC which can push the parameter space into the alignment limit.

Recently, there has been increasing interest in placing detectors far from the interaction point to detect weakly interacting BSM particles. When $H_5^{\pm\pm}$ leaves the detector, it may also induce signals at various far detectors. In the analysis, we considered MATHUSLA, ANUBIS, FACET, FASER, CODEX-b and MoEDAL-MAPP and AL3X. We found that the exclusion area is influenced by the combined effect of the relative position of far detectors and IP, the size of the far detectors, the integrated luminosity, the decay time of detected particles and the cross-section. For the very forward facilities(FACET and FASER), the sensitivity is weak due to the small acceptance area of these detectors. In HL-LHC, the far detectors around the central region, MATHUSLA, ANUBIS, CODEX-b and MoEDAL-MAPP, the coverage extends from several meters to $\mathcal{O}(10^4)$ meters.

Note that, in current analysis, we didn't consider the case where $H_5^{\pm\pm}$ decays within the calorimeters, which requires a dedicated simulation. However, from~\autoref{fig:LLP_Exclusion}, we find that, beside some small gap in the low mass region, the searches from inner tracker, muon system and HSCP can already cover the whole parameter region that leads to a long-lived $H_5^{\pm\pm}$ in the GM model.

Searches for long-lived particles play an important role in covering the parameter space of the GM model. The usual direct searches generally cover parameter space with large $s_H$ (or equivalently large triplet vev $v_\chi$), the model can always escape the stringent constraints by pushing to the alignment limit ($s_H\to 0$). However, the searches for long-lived particles alter the situation. It can cover the parameter space around the alignment limit and hence will be a powerful complimentarity to the usual searches covering the whole parameter space of the GM model.

\begin{acknowledgments}
We would like to thank Zeren Simon Wang for helpful discussion on details of DDC. C.-T. Lu and X. Wei are supported the National Natural Science Foundation of China (NNSFC) under grant No.~12335005 and the Special funds for postdoctoral overseas recruitment, Ministry of Education of China. X. Wang and Y. Wu are supported by the NNSFC under grant No.~12305112.
\end{acknowledgments}

\appendix

\section{Trajectories of particles affected by magnetic fields in the LHC detectors}\label{appendix:trajectory}
Given a particle with momentum $(E,p_x,p_y,p_z)$, charge $q$ and a magnetic field aligned along $z$-axis, the motion of such particle can be described as a combination of two movements: movement with constant velocity along $z$-axis and circular movement in the transverse $x$-$y$ plane. Assuming that the strength of the magnetic field is $B$, the radius of the trajectory in the transverse plane is given by
\begin{align}
    R = \frac{p_T}{qB},
\end{align}
where $p_T = \sqrt{p_x^2 +p_y^2}$. Then the coordinates of the charged particle as functions of time are given by
\begin{align}
    x &= \frac{2p_T}{qB}\sin\left(\frac{qBt}{2E}\right)\sin\left(\frac{qBt}{2E}+\frac{\pi}{2}-\theta_0\right), \\
    y &= \frac{2p_T}{qB}\sin\left(\frac{qBt}{2E}\right)\cos\left(\frac{qBt}{2E}+\frac{\pi}{2}-\theta_0\right), \\
    z &= \frac{p_z}{E} t,
\end{align}
where $\theta_0$ is the angle between the momentum and the $x$-axis in the transverse plane at the beginning.

\bibliographystyle{bibsty}
\bibliography{references}

\end{document}